\documentclass[review]{elsarticle}

\usepackage{moreverb,url}

\usepackage[colorlinks,bookmarksopen,bookmarksnumbered,citecolor=red,urlcolor=red]{hyperref}

\usepackage{amsmath}
\usepackage{amsthm}
\usepackage{amssymb}
\usepackage{amsbsy}
\usepackage{mathrsfs}
\usepackage{graphicx}
\usepackage{epsfig}
\usepackage{bm}
\usepackage{natbib}
\usepackage{multirow} 
\usepackage{longtable}

\def\c#1{\ensuremath{\mathcal{#1}}}

\newcommand{\nmathbf}{\bm}

\def\bfY{\nmathbf Y}

\def\bfu{\nmathbf u}

\def\bfx{\nmathbf x}
\def\bfy{\nmathbf y}

\def\bfbeta   {\nmathbf \beta}

\def\bftheta  {\nmathbf \theta}

\newcommand{\cfB}{\mbox{\c{B}}}

\newcommand{\cfS}{\mbox{\c{S}}}

\def\boldfacefake#1{\kern-4pt
   \hbox{ \mathsurround=0pt
   \hbox to 0.4pt{$#1$\hss}\hbox to 0.4pt{$#1$\hss}\hbox {$#1$}}}

\newcommand{\E}{\mbox{E}}

\newcommand{\Var}{\mbox{Var}}

\newcommand{\btable}{\begin{table}[h]\centering}
\newcommand{\etable}{\end{table}}
\newcommand{\bt}{\begin{parag}\small \let\b=\nsb \let\sb=\nssb \begin{tabular}}
\newcommand{\et}{\end{tabular}\let\b=\nb \let\sb=\nsb\end{parag}}

\newenvironment{parag}{\par}{\par}

\newcommand{\be}{\begin{eqnarray}}
\newcommand{\ee}{\end{eqnarray}}
\newcommand{\ba}{\begin{eqnarray*}}
\newcommand{\ea}{\end{eqnarray*}}

\newcommand{\reals}{\mbox{\rm I\kern-.20em R}}
\newcommand{\sreals}{\mbox{\small \rm I\kern-.20em R}}

\begin{document}

%\runninghead{Marginalized Beta ITS}

\title{A marginalized three-part interrupted time series regression model for proportional data}

\author[1]{Shangyuan Ye\corref{cor1}}
\ead{yesh@ohsu.edu}

\author[2]{Maricela Cruz}
\ead{Maricela.F.Cruz@kp.org}

\author[3]{Ziyou Wang}
\ead{ziyou.wang@kcl.ac.uk}

\author[1]{Yun Yu}
\ead{yuy@ohsu.edu}

\cortext[cor1]{Corresponding author}
\affiliation[1]{organization={Biostatistics Shared Resource, Knight Cancer Institute, Oregon Health \& Science University}}

\affiliation[2]{organization={Kaiser Permanente Washington Health Research Institute and Department of Biostatistics, School of Public Health, University of Washington}}

\affiliation[3]{organization={Department of Mathematics, King's College London}}

\begin{abstract}
Interrupted time series (ITS) is often used to evaluate the effectiveness of a health policy intervention that accounts for the temporal dependence of outcomes. When the outcome of interest is a percentage or percentile, the data can be highly skewed, bounded in $[0, 1]$, and have many zeros or ones. A three-part Beta regression model is commonly used to separate zeros, ones, and positive values explicitly by three submodels. However, incorporating temporal dependence into the three-part Beta regression model is challenging. In this article, we propose a marginalized zero-one-inflated Beta time series model that captures the temporal dependence of outcomes through copula and allows investigators to examine covariate effects on the marginal mean. We investigate its practical performance using simulation studies and apply the model to a real ITS study.
\end{abstract}

\maketitle

\section{Introduction}
Interrupted time series (ITS) design, arguably the most powerful quasi-experimental design \citep{cook2002experimental}, is often used to evaluate the effectiveness of a health policy intervention that accounts for the temporal dependence of outcomes \citep{wagner2002segmented,penfold2013use,bernal2017interrupted}. Aggregated-level outcomes are repeatedly collected before and after policy intervention in ITS designs \citep{kontopantelis2015regression}, and segmented time series regression is the most popular method for analysis \citep{cook2002experimental,cruz2017robust}.

Percentages or percentiles are commonly used outcomes in ITS designs and are typically analyzed using linear segmented time series regression (by assuming the percentages are normally distributed) \citep{van2009influence}. However, as mentioned in \cite{ferrari2004beta,chai2018marginalized}, linear models are not appropriate for proportional data because (1) estimates of the regression parameters can exceed their lower and upper bounds, and (2) data can be highly skewed and have many zeros or ones, which violates the normality assumption. 

One approach for analyzing proportional data is to first apply a logistic transformation on the response variable and then assume a linear regression model on the transformed response variable. Under scenarios where there are many zeros or ones, the three-part logistic transformation model proposed by \cite{fang2013three} can be used. However, one drawback of this approach is that the regression parameters can not be directly interpreted on the original response scale as a consequence of Jensen's inequality \citep{guolo2014beta,kieschnick2003regression}.

Beta regression, proposed by \cite{ferrari2004beta}, provides an alternative method of analyzing proportional data. Extensions include zero-or-one inflated \citep{ospina2012general}, zero-one inflated \citep{abdel2017extended}, and marginalized zero-inflated \citep{chai2018marginalized} Beta regressions, all dealing with outcomes with many zeros or ones. 

To the best of our knowledge, there are no existing models for serially dependent zero-one inflated proportional data. To fill this gap, we propose a marginalized zero-one inflated Beta regression time series model for analyzing zero-one inflated proportional outcomes in ITS designs. This work is motivated by data from a study aimed to assess the impact of a new care delivery model on patient experience survey scores for a single hospital, tracked monthly between January 2008 and December 2012 \citep{bender2015clinical}. Patient experience scores are nationally endorsed quality and safety metrics used to calculate health systems’ reimbursement for care services via the Center for Medicaid and Medicare Services value-based purchasing program, thus making the patient experience scores a focus for improvement \citep{kavanagh}. There are several patient experience indicators, including `nurse communication', `skill of the nurse', and `pain management', but for the purposes of showcasing our methodology, we focus on the score on `pain management'. The intervention, administered in July of 2010, was the implementation of Clinical Nurse Leader integrated care delivery (CNL), a new nursing care delivery model, which embedded a master-prepared nurse with advanced competencies in clinical leadership, care environment management, and clinical outcomes management into the front lines of care \citep{Bender2017}. The nurses were introduced into their respective hospital units in January 2010, six months prior to the formal intervention implementation time, while conducting their master's level microsystem change project. This early introduction had the ability to influence the ‘change point’ of the intervention effect. Prior studies have assessed the impact of nursing care delivery interventions on patient experience scores via ITS methods, but these studies assumed the patient experience scores were normally distributed and did not allow for anticipated or delayed intervention effects \citep{bender2019system,Bender:2012ch}. 

Existing time series models for non-Gaussian data can be classified into three categories: observation-driven models, parameter-driven models, and copula-based models. In observation-driven models, the correlation of outcomes is specified through the direct incorporation of lagged values of the observed data into the mean function of the model. Examples include generalized linear autoregressive moving average models \citep{davis2003observation} or log-linear models \citep{fokianos2009poisson} for count outcomes and Beta autoregressive processes \citep{casarin2012bayesian} for proportional data (without inflation). Observation-driven models are appealing in prediction because of the straightforward likelihood inference. However, interpreting regression parameters can be challenging because these parameters represent the effects conditionally on past observations. Parameter-driven models \citep{da2016hierarchical,sorensen2019independence}  specify the correlation of outcomes through a latent process. Although parameter-driven models are attractive because they share the same parameter interpretation as generalized linear models for independent data, parameter estimation is more challenging due to the latent process. 

As an alternative to the observation- and parameter-driven models, copula-based approaches separately specify the marginal model and dependence structure according to Sklar's theorem \citep{joe1997multivariate}. \cite{masarotto2012gaussian} introduced Gaussian copula marginal regression models and included count time series data as an example. \cite{guolo2014beta} proposed a Gaussian copula-based Beta regression model to analyze the influenza-like-illness incidence data. \cite{alqawba2021copula} proposed a family of copula-based Markov zero-inflated count time series models to analyze the airport sandstorm data. 

In this article, we consider the copula-based approach to construct the time series through copula-based joint distributions of consecutive observations by proposing a model, with interpretable regression parameters, that accounts for zero-one-inflation in proportional data time series. The paper is organized as follows: Section \ref{sec:M} introduces the proposed model and Section \ref{sec:inference} describes the details on the estimation and inference procedure. Section \ref{sec:simulation} evaluates the finite sample performance of the proposed estimation procedure and Section \ref{sec:app} uses the proposed method to assess the impact of a new delivery model on patient experience survey scores. We end the paper with a discussion in Section \ref{sec:con}.

\section{Marginalized zero-one-inflated Beta time series models} \label{sec:M}
\subsection{Marginalized zero-one-inflated Beta regression models}\label{subsec:MB}
We first introduce the marginalized zero-one-inflated Beta regression models (MZOIB). For $t = 1, \cdots, n$, let $Y_t \in [0, 1]$ and $\bfx_t\in\mathcal{X}$ be the outcome and observed covariates at time $t$, respectively. We define the latent binary variable $d_{1t}$ to be a binary variable that indicates whether the response is nonzero, and $d_{2t}$ to be a binary variable that indicates whether the response is equal to one, i.e., $d_{1t}=I(Y_t > 0)$ and $d_{2t}=I(Y_t = 1)$. Let $p_{1t} = P(d_{1t} = 1)$ and $p_{2t} = P(d_{2t} = 1 |  d_{1t} = 1)$. We further assume that the random variable $Y_t,$ conditional on $Y_t \in (0, 1),$ follows a Beta distribution using the parameterization specified by \cite{ferrari2004beta}, in which the probability density function (PDF) can be written as
\be \label{beta}
f_{t}(y_t ~|~ 0<Y_t<1) := g(y_t; \mu_t, \phi_t) = \frac{\Gamma(\phi_t)}{\Gamma(\mu_t \phi_t) \Gamma[(1-\mu_t) \phi_t]} y_t^{\mu_t\phi_t - 1} (1-y_t)^{(1-\mu_t)\phi_t - 1},
\ee
where $\mu_t \in (0, 1)$ and $\phi_t > 0$ are the mean and dispersion parameters, respectively. The mean and variance of $Y_t$ conditional on $Y_t \in (0, 1)$ can be expressed as
\be \label{BMV}
\E(Y_t ~|~ 0<Y_t<1) = \mu_t ~~\text{and}~~ \Var(Y_t ~|~ 0<Y_t<1) = \frac{\mu_t (1-\mu_t)}{1 + \phi_t}.
\ee
Then, we say the unconditional distribution of $Y_t$ is a Zero-One-Inflated Beta distribution with parameters $p_{1t}, p_{2t}, \mu_t,$ and $\phi_t$, i.e., $Y_t \sim ZOIB(p_{1t}, p_{2t}, \mu_t, \phi_t)$. The PDF of $Y_t$ can therefore be expressed as
\be \label{density}
f_{t}(y_t; p_{1t}, p_{2t}, \mu_t, \phi_t) &=& (1-p_{1t}) I(y_t = 0) + p_{1t}[p_{2t} I(y_t = 1) \nonumber \\ &~& + (1 - p_{2t}) g(y_t; \mu_t, \phi_t) I(0 < y_t < 1)].
\ee
The mean and variance of $Y_t$ are
\be \label{moment}
\E(Y_t) &=& v_t = p_{1t} [(1-p_{2t}) \mu_t + p_{2t}] \nonumber \\ 
\Var(Y_t) &=& p_{1t} (1-p_{2t}) \left\{ \frac{\mu_t (1-\mu_t)}{1 + \phi_t} + [1 - p_{1t} (1-p_{2t})] \mu_t^2 + 2 p_{1t} p_{2t} \mu_t \right\} \nonumber\\
&&\qquad\qquad\qquad+ p_{1t} p_{2t} (1- p_{1t} p_{2t}). 
\ee
Via equation (\ref{moment}), we then obtain  
\be \label{CM}
\mu_t = \frac{1}{1-p_{2t}} \left( \frac{v_t}{p_{1t}} - p_{2t} \right).
\ee
When $p_{2t} = 0$ or $p_{1t} = 1$, equation (\ref{density}) reduces to the PDFs of the zero-inflated or one-inflated Beta, respectively. 

To model the relationship between covariates and the outcome, we first assume two separate logistic regression models on the binary latent variables $d_{1t}$ and $d_{2t}$, i.e.
\be \label{binaryP}
\log \left( \frac{p_{1t}}{1 - p_{1t}} \right) = \bfx_{1t}^\top \bfbeta_1 ~~ \text{and} ~~ \log \left( \frac{p_{2t}}{1 - p_{2t}} \right) = \bfx_{2t}^\top \bfbeta_2,  
\ee
where $\bfx_{1t}, \bfx_{2t}$ are subvectors of $\bfx_t$, and $\bfbeta_1$ and $\bfbeta_2$ are vectors of corresponding coefficients. To obtain interpretable regression parameters on the marginal (unconditional) mean, we consider the following marginalized model, which generalizes the marginalized two-part beta regression model proposed by \cite{chai2018marginalized}. Since $v_t \in (0, 1)$, we consider the following logistic regression model
\be \label{MM}
\text{logit}(v_t) =\log \left( \frac{v_t}{1-v_t} \right) = \bfx_{3t}^\top \bfbeta_3,
\ee
where $\bfx_{3t}$ is a subvector of $\bfx_t$ and $\bfbeta_3$ is the vector of corresponding coefficients. Additionally, we assume a log-linear model on the dispersion parameter $\phi_t$ to allow its value to depend additionally on covariates:
\ba
\log(\phi_t) = \bfx_{4t}^\top \bfbeta_4
\ea
where $\bfx_{4t}$ is a subvector of $\bfx_t$ and $\bfbeta_4$ is the vector of corresponding coefficients.

\subsection{Copula-based MZOIB time series models} \label{subsec:MBTS}
In this paper, we propose a copula-based MZOIB time series model (MZOIBTS). We assume the sequence of random variables $\{ Y_t \}_{t=1}^T$ follows a first-order Markov chain, i.e.,
\be \label{MC1}
Y_t = h(\epsilon_t; Y_{t-1}),
\ee
where $\{ \epsilon_t \}$ is an independent identically distributed (i.i.d.) stochastic latent process with $\E(\epsilon_t) = 0$ and $h(\cdot)$ is an increasing function \citep{joe2014dependence}. Therefore, the joint PDF of $\bfY$ can be decomposed as
\be \label{joint}
f(\bfy; \bftheta, \rho) = f_1(y_1; \bftheta) \prod_{t=2}^n f(y_t | y_{t-1}; \bftheta, \rho) = f_1(y_1; \bftheta) \prod_{t=2}^n \frac{f_{t,t-1}(y_t, y_{t-1}; \bftheta, \rho)}{f_{t-1} (y_{t-1}; \bftheta)},
\ee
where $\bfy = (y_1, \cdots, y_n)^\top$ and $\rho$ is a parameter that measures the dependence between $Y_t$ and $Y_{t-1}$. Because there is no multivariate extension of the MZOIB density (\ref{density}), we proposed to use copulas to construct the joint distribution $f_{t,t-1}(y_t, y_{t-1}; \bftheta, \rho)$.

We give a brief introduction to copulas in this paper and refer the readers to \cite{joe1997multivariate,nelsen2007introduction,joe2014dependence} for further discussion on theorems and extensions. For a $d$-dimensional random variable $\bfY = (Y_1, \cdots, Y_d)^\top$, according to Sklar's theorem \citep{nelsen2007introduction}, the copula function $C: [0, 1]^d \rightarrow [0, 1]$ of $\bfY$ is defined as
\ba
C(u_1, \cdots, u_d) = H(F_1^{-1}(u_1), \cdots, F_d^{-1}(u_d))
\ea
where $H(y_1, \cdots, y_d) = P(Y_1 \le y_1, \cdots, Y_d \le y_d)$ is the joint cumulative distribution function (CDF) of $\bfY$, $F_t(\cdot)$ is the marginal CDF of $Y_t$, $F_t^{-1}(\cdot)$ is the inverse function of $F_t(\cdot)$, and $u_t \in [0, 1]$ for $t = 1, \cdots, d$. Conversely, given the copula function  of $\bfY$, its CDF can be re-expressed as
\be \label{copula}
H(y_1, \cdots, y_d) = C(F_1(y_1), \cdots, F_d(y_d)).
\ee

For any $u_t \in [0, 1]$, $t = 1, \cdots, n$, a valid copula function $C$ should satisfy the following three conditions:
\begin{enumerate}
    \item $C(1, \cdots, u_t, \cdots, 1) = u_t$;
    \item $C(u_1, \cdots, u_n) = 0$ if $\prod_{t=1}^nu_t = 0$;
    \item For any $\cfB = \prod_{t=1}^d [u_{t1}, u_{t2}] \in [0, 1]^d$, there is
    \be \label{integral}
    \int_{\cfB} dC(\bfu) = \sum_{j_1 = 1}^2 \cdots \sum_{j_d = 1}^2 (-1)^{\sum_{t=1}^d j_t} C(u_{1j_1}, \cdots, u_{dj_d}) \ge 0.
    \ee
\end{enumerate}
A wide range of parametric copula families satisfying the above properties have been proposed and studied \citep{nelsen2007introduction}. Table \ref{Copula function} summarizes some commonly used copula functions that we will consider in this paper. 

We then complete the model of MZOIBTS through a bivariate copula function. Under (\ref{copula}), we have $H(y_t, y_{t-1}; \bftheta, \rho) = C(F_t(y_t; \bftheta), F_{t-1}(y_{t-1}; \bftheta); \rho)$. When both $y_t, y_{t-1} \in (0, 1)$, under the ZOIB model (\ref{density}), the mapping between $y_t, y_{t-1}$ and $F_t(y_t; \bftheta), F_{t-1}(y_{t-1}; \bftheta)$ is one-to-one. Thus, we can write
\be \label{joint_cc}
f_{t,t-1}(y_t, y_{t-1}; \bftheta, \rho) = f_t(y_t; \bftheta) f_{t-1}(y_{t-1}; \bftheta) c(u_t, u_{t-1}; \rho), 
\ee
where $u_t = F_t(y_t; \bftheta)$, $u_{t-1} = F_{t-1}(y_{t-1}; \bftheta)$, and $c(u_t, u_{t-1}; \rho) = \frac{\partial^2}{\partial u_t \partial u_{t-1}} C(u_t, u_{t-1}; \rho)$ is the copula density function. On the other hand, if either or both of $y_{t-1}$ and $y_t$ are 0 or 1, the mapping between $y_t, y_{t-1}$ and $F_t(y_t; \bftheta), F_{t-1}(y_{t-1}; \bftheta)$ would then be many-to-one. Define $u_t = (1-p_{1t})I(y_t = 0)+I(y_t = 1)$ and $u_t^{-} = (1 - p_{1t} p_{2t})I(y_t = 1)$. By the third property (equation (\ref{integral})) of the copula functions, when both $y_{t-1}$ and $y_t$ equal 0 or 1, we have
\be \label{joint_dd}
P(Y_t = y_t, Y_{t-1} = y_{t-1}; \bftheta, \rho) &=& C(u_t, u_{t-1}; \rho) - C(u_t^{-}, u_{t-1}; \rho) - C(u_t, u_{t-1}^{-}; \rho) \nonumber \\ &+& C(u_t^{-}, u_{t-1}^{-}; \rho);
\ee
when $y_t \in \{ 0, 1 \}$ and $y_{t-1} \in (0, 1)$, we have
\be \label{joint_dc}
f(Y_t = y_t, y_{t-1}; \bftheta, \rho) = \frac{\partial}{\partial u_{t-1}} [C(u_t, u_{t-1}; \rho) - C(u_t^{-}, u_{t-1}; \rho)] f_{t-1} (y_{t-1}; \bftheta);
\ee
finally, when $y_t \in (0, 1)$ and $y_{t-1} \in \{ 0, 1 \}$, we have
\be \label{joint_cd}
f(y_t, Y_{t-1} = y_{t-1}; \bftheta, \rho) = \frac{\partial}{\partial u_{t}} [C(u_t, u_{t-1}; \rho) - C(u_t, u_{t-1}^{-}; \rho)] f_{t} (y_{t}; \bftheta). 
\ee

\subsection{Interrupted time series analysis for MZOIBTS} \label{subsec:ITS}
Segmented time series regression is the most commonly used method for analyzing data of ITS studies \citep{penfold2013use,cruz2017robust}. For a single-arm ITS design \citep{wagner2002segmented,ye2022comparison} with zero-one inflated proportional outcomes, we consider the following generalized segmented linear regression model
\be \label{segment}
\text{logit}(v_t) = \bfx_{3t}^\top \bfbeta_3 = \beta_{30} + \beta_{31} T(t) + \beta_{32} I(T(t) \ge T(\tau)) + \beta_{33} (T(t)-T(\tau))_+,
\ee
where $T(\cdot)$ is an increasing function in $t$ (e.g., $T(t) = \log(t)$ or $T(t) = t$), $\tau$ is the time point when the intervention is initiated (which is often called ``change point''), $\bfx_{3t} = (1, T(t), X_t, T(t)-T(\tau)_+)^\top$, and $\bfbeta_3 = (\beta_{30}, \beta_{31}, \beta_{32}, \beta_{33})^\top$ denotes the corresponding regression parameters, with $\beta_{30}$ representing the starting level of the logit-transformed marginal mean, $\beta_{31}$ representing the slope of the logit-transformed marginal mean before the change point ($\tau$), $\beta_{32}$ representing the immediate change of the logit-transformed marginal mean after the intervention, and $\beta_{33}$ representing the difference in the slopes of the logit-transformed marginal mean after the intervention. Denote $t_0$ as the time point of policy intervention, although \cite{penfold2013use,ye2022comparison,rhee2021decline}; and many others, assumed an instantaneous effect after intervention ($t_0 = \tau$), it is also possible that the time of change point differs from the time of policy intervention (i.e., either $t_0 > \tau$ or $t_0 < \tau$) \citep{cruz2017robust,cruz2019assessing}. Thus, we assume $\tau$ is unknown and needs to be estimated based on the observed data.

For the parts of the model that deal with excess zeros and ones (i.e., model (\ref{binaryP})), we only include intercepts due to the small sample sizes typical in ITS studies.

\section{Statistical inference} \label{sec:inference}
Exact likelihood inference for the proposed MZOIBTS model is computationally challenging because the likelihood surface is ill-behaved due to the marginal CDF transformation in the copula function \citep{joe2005asymptotic}. Thus, we propose to estimate the marginal model parameters through the composite marginal likelihood \citep{varin2011overview} and consider two methods for standard error estimation. The first method corrects the model misspecification through the heteroskedasticity and autocorrelation consistent (HAC) covariance estimation procedure \citep{newey1986simple}. The second method utilizes the pseudo maximum likelihood estimator for copula parameter estimation and estimates the standard errors through parametric bootstrap. The procedure is a special case of the inference function for margin (IFM) approach proposed by \cite{joe1996estimation}.  

\subsection{Estimation of marginal parameters} \label{subsec:stage1}
The main interest in ITS analysis is to make inference on the marginal parameters, $\bftheta$. We propose to estimate $\bftheta$ by maximizing the composite log-likelihood function under the assumption of independence, as suggested by \cite{chandler2007inference}:  
\be
cl(\bftheta; \bfy) &=& \sum_{t=1}^n \log f(y_t; \bftheta) = \sum_{t=1}^n l_t(\bftheta; y_t) \\ &=& -\sum_{t=1}^n \log(1 + \exp(\bfx_{1t}^\top \bfbeta_1)) + \sum_{t: y_t>0} \{ \bfx_{1t}^\top \bfbeta_1 - \log(1 + \exp(\bfx_{2t}^\top \bfbeta_2)) \} \nonumber \\ &~& + \sum_{t: 0<y_t<1} \left\{ h_t + \exp(\bfx_{4t}^\top \bfbeta_4) \log(1-y_t) - \log(y_t (1-y_t)) \right. \nonumber \\ &~& + \left. \mu_t \exp(\bfx_{4t}^\top \bfbeta_4) \text{logit}(y_t) \right\} + \sum_{t: y_t=1} \bfx_{2t}^\top \bfbeta_2, \nonumber
\ee
where
\ba
h_t = h_t(\bfbeta_1, \bfbeta_2, \bfbeta_3, \bfbeta_4) &=& \log \Gamma (\exp(\bfx_{4t}^\top \bfbeta_4)) - \log \Gamma (\mu_t \exp(\bfx_{4t}^\top \bfbeta_4)) \\ &~& - \log \Gamma ((1 - \mu_t) \exp(\bfx_{4t}^\top \bfbeta_4)), \\
\mu_t = \mu_t(\bfbeta_1, \bfbeta_2, \bfbeta_3) &=& \frac{(1 + \exp(\bfx_{2t}^\top \bfbeta_2)) (1 + \exp(-\bfx_{1t}^\top \bfbeta_1))}{1 + \exp(-\bfx_{3t}^\top \bfbeta_3)} - \exp(\bfx_{2t}^\top \bfbeta_2).
\ea
This is known as the pseudo maximum likelihood estimation procedure under working independence assumptions, sometimes also referred to as the composite marginal likelihood estimation \citep{varin2011overview}. The corresponding composite score function, $u_t(\bftheta) = \nabla l_t(\bftheta; y_t)$ is detailed in the Appendix. The composite score function can be used for standard error estimations and model selection, which will be elaborated in the following subsections.

\subsection{Standard error estimation} \label{subsec:stage2}
Although the composite likelihood estimator $\hat{\bftheta}$ introduced in Section \ref{subsec:stage1} already provides a consistent estimation of the regression parameters of interest, valid inference on these parameters (e.g., standard errors, confidence intervals, or hypothesis tests) requires extra consideration on the autocorrelation between outcomes.

\subsubsection{HAC covariance estimation} \label{subsec:HAC}
Under certain regularity conditions, it has been shown that the composite likelihood estimator $\hat{\bftheta}$ is consistent and asymptotically normally distributed with
\ba
\sqrt{n} (\hat{\bftheta} - \bftheta) \xrightarrow{d} N(\nmathbf{0}, G^{-1}(\bftheta)),
\ea
where $G(\bftheta) = H(\bftheta) J^{-1}(\bftheta) H(\bftheta)$ is the Godambe information matrix \citep{godambe1960optimum}, with
$H(\bftheta) = \E\{ -\nabla^2 cl(\bftheta; \bfY) \}$ being the sensitivity matrix and $J(\bftheta) = \Var(\nabla cl(\bftheta; \bfY))$ being the variability matrix. Although, the sensitivity matrix can be consistently estimated by the negative Hessian matrix of the composite likelihood evaluated at the $\hat{\bftheta}$u under standard regularity conditions:
\ba
\hat{H}(\bftheta) = -\nabla^2 cl(\hat{\bftheta}; \bfy) = -\sum_{t=1}^n \nabla u_t (\hat{\bftheta}), 
\ea
difficulties arise for the estimation of the variability matrix due to the heteroskedasticity and autocorrelation of the MZOIBTS model. Here, we consider the HAC covariance estimation procedure proposed by \cite{newey1986simple}, where the variability matrix is estimated by the average of weighted sums of lagged products of the estimated score functions:
\ba
\hat{J}(\bftheta) = \frac{1}{n} \sum_{t=1}^n \sum_{t'=1}^n w_{|t-t'|} u_{t} (\hat{\bftheta}) u_{t'}^\top (\hat{\bftheta}),
\ea
where $\{w_l\}$ is a sequence of linearly decay weights defined as
\ba
w_l = \max \left( 0, 1 - \frac{l}{L+1} \right),
\ea
where $L$ is the maximum lag.

\subsubsection{Parametric bootstrap estimator} \label{subsec:SE}
An alternative approach is to first estimate the copula parameter and then estimate the standard errors of $\hat{\bftheta}$ via parametric bootstrap. Here, we consider the pseudo maximum likelihood copula estimator \citep{gong1981pseudo}, where the copula parameter is estimated by maximizing the likelihood function conditioning on the estimated parameters in the first stage, i.e.,  
\ba
\hat{\rho} = \arg\max_\rho L(\rho; \bfy, \hat{\bftheta}) = \arg\max_\rho \prod_{t=2}^n f_{t,t-1}(y_t, y_{t-1}; \hat{\bftheta}, \rho) = \arg\max_\rho \prod_{t=2}^n \tilde{C}(\hat{u}_t, \hat{u}_{t-1}; \rho),
\ea
where $\hat{u}_t = F_t(y_t; \hat{\bftheta})$, $\hat{u}_{t-1} = F_{t-1}(y_{t-1}; \hat{\bftheta})$, and
\ba
\tilde{C}(\hat{u}_t, \hat{u}_{t-1}; \rho) = \begin{cases} c(\hat{u}_t, \hat{u}_{t-1}; \rho) & \text{ if } y_t, y_{t-1} \in (0, 1) \\ P(Y_t = y_t, Y_{t-1} = y_{t-1}; \hat{\bftheta}, \rho) & \text{ if } y_t, y_{t-1} \in \{0, 1\} \\ \frac{\partial}{\partial u_{t-1}} [C(\hat{u}_t, \hat{u}_{t-1}; \rho) - C(\hat{u}_t^{-}, \hat{u}_{t-1}; \rho)] & \text{ if } y_t \in \{0, 1\}, y_{t-1} \in (0, 1) \\ \frac{\partial}{\partial u_{t}} [C(\hat{u}_t, \hat{u}_{t-1}; \rho) - C(\hat{u}_t, \hat{u}_{t-1}^{-}; \rho)] & \text{ if } y_t \in (0, 1), y_{t-1} \in \{0, 1\}. \end{cases}
\ea

Conditioning on the composite likelihood estimator of $\bftheta$ and the pseudo maximum likelihood estimator of $\rho$, $R$ trajectories of $\bfY$ are simulated through the MZOIBTS model with parameters $(\bftheta, \rho) = (\hat{\bftheta}, \hat{\rho})$ and covariates $\bfx$ as in the original data. For each bootstrapped dataset, we calculate the composite likelihood estimator, denoted as $\hat{\bftheta}^r$, as introduced in Section \ref{subsec:stage1}. The standard errors of $\hat\bftheta$ are then estimated as the sampling standard deviations of the bootstrap estimates, i.e., $\hat{\text{se}}(\hat{\bftheta}) = \text{sd}(\hat{\bftheta}^r)$.

\subsection{Confidence interval and hypothesis testing}
Because $\hat{\bftheta}$ is consistent and asymptotically normally distributed, we suggest to construct the $(1 - \alpha)*100$\% confidence interval for $\theta_j$ with $\hat{\theta}_j \pm z_{\alpha/2} \hat{\text{se}}(\hat{\theta}_j)$, where $z_{\alpha/2}$ is the $(1-\alpha/2)^{th}$ quantile of the standard normal distribution and $\hat{\text{se}}(\hat{\theta}_j)$ is one of the standard error estimators introduced in Section \ref{subsec:stage2}. We use the confidence interval based on a normal approximation as it usually requires a smaller number of bootstrap datasets $R$ than the confidence interval based on the quantiles of $\hat\bftheta^r$ \citep{davison1997bootstrap}.

For hypothesis tests of the form $H_0: A\bftheta = \mathbf{0}$ versus $H_1: A\bftheta \ne \mathbf{0}$, where $A$ is a fixed matrix, we consider the Wald-type test statistic, which rejects the null hypothesis if
\ba
W = [A \hat{\bftheta}]^\top [A \hat{V}(\hat{\bftheta}) A^\top]^{-1} [A \hat{\bftheta}] > \chi^2_{Q} (1-\alpha)
\ea
where $\alpha$ is the significance level, $\hat{V}(\hat{\bftheta})$ is the estimator of the covariance matrix of $\hat{\bftheta}$, and $\chi^2_{Q}(\cdot)$ is the quantile function of the chi-square distribution with degrees of freedom, $Q,$ equal to the rank of matrix $A$. We use a Wald-type test over the independence likelihood ratio test statistic because the latter is not recommended here due to its non-standard asymptotic distribution \citep{varin2011overview}.

\subsubsection*{Hypothesis tests for ITS analysis}
%\begin{remark}[Hypothesis tests for ITS analysis]
For ITS analysis, we are typically interested in testing whether or not there exists (1) a level change, (2) a trend change, or (3) any changes after policy intervention. From model (\ref{segment}), the corresponding tests are: (1) $H_0: \beta_{32} = 0$ versus $H_1: \beta_{32} \ne 0$, which tests for a level change after intervention; (2) $H_0: \beta_{33} = 0$ versus $H_1: \beta_{33} \ne 0$, which tests the trend change after intervention; and (3) $H_0: \beta_{32} = \beta_{33} = 0$ versus $H_1$: any of the $\beta_{3j} \ne 0$ for $j = 2$ or 3, which tests if there exist any changes (level, trend, or both) after intervention.
%\end{remark}

\subsection{Model selection} \label{subsec:selection}
In ITS analysis, it is possible that the change point $\tau$ is not the same as the time of policy intervention $t_0$. In this case, one option is to utilize variable selection measures, such as the composite likelihood-based Akaike (AIC) and Bayesian (BIC) information criteria derived by \cite{varin2005note} and \cite{gao2010composite}, respectively, to estimate $\tau$. Specifically, let $\cfS^{\tau}$ be a set of possible values of the change point, $\tau,$ obtained from experts. The composite likelihood BIC, denoted as cBIC, selects the optimal change by minimizing the following objective function:
\ba
\mathrm{cBIC} = -l_{ind}^{\tau}(\bftheta; \bfy) + \log(n) \mathrm{tr} \left\{ [J^{\tau}(\bftheta)]^{-1} H^\tau(\bftheta) \right\}.
\ea
Here we write the independence log-likelihood, as well as the sensitivity and variability matrices, as $l_{ind}^{\tau}(\bftheta; \bfy)$, $J^{\tau}(\bftheta)$, and $H^\tau(\bftheta)$, respectively, to emphasize that its value depends on the change point $\tau$. 

Similarly, we can also select the best approximating parametric copula family by maximizing the pseudo likelihood introduced in Section \ref{subsec:stage2}. Because fitting the pseudo likelihood can be time-consuming for some copula functions (e.g. Clayton copula), we suggest using Gaussian copula as the default model since previous studies have shown its robustness against model misspecification \citep{masarotto2012gaussian}. 

\section{Simulation studies} \label{sec:simulation}
To evaluate the finite sample performance of the proposed two-step estimators in interrupted time series analyses, we conduct simulation studies with outcomes generated from the model specified in Section \ref{sec:M} with Gaussian and Frank copulas. We assume intercept-only models for the zero-part and one-part models. We set $\beta_1 = 2.944$ and $\beta_2 = -2.197$, corresponding to $p_1 = 0.95$ and $p_2 = 0.1$ respectively. For the marginal mean model, we consider the generalized segmented linear regression model (\ref{segment}), with the true regression parameter set to be $\bfbeta_3 = (0.847, -0.01, -0.5, -0.3)^\top$ and $T(t) = \log(t)$. We allocate equal numbers of observations uniformly before and after the intervention, i.e., $t_0 = (n+1)/2$. The dependence parameter for the Gaussian copula is chosen as $\rho = 0.5, 0.2, -0.2, -0.5$, while the dependence parameter for the Frank copula was chosen as $\rho = 3.3, 1.17, -1.17, -3.3$, corresponding to the same level of dependencies as the ones specified for Gaussian copula. Because we observe a significant dispersion change in the real data analysis (see Section \ref{sec:app}), we fix $\phi_t = 20$ ($\beta_{40} = 3$) for $t \le \tau$ and $\phi_t = 33$ ($\beta_{41} = 0.5$) for $t > \tau$. We simulate $K = 2,000$ time series for each setting to evaluate the empirical performance. Both HAC and bootstrap estimators introduced in Section \ref{sec:inference} are considered for standard error estimation. This allows us to evaluate and compare the performance of the proposed estimators under different conditions and identify which method provides more reliable standard error estimates in the context of our simulation study.

Because for ITS designs, we are usually interested in making inferences about the level change $\beta_{32}$ and the slope change $\beta_{33}$, our comparison mainly focuses on these two parameters. Bias, standard error of $\hat{\beta}$ ($\mbox{SE}(\hat{\beta})$), mean of standard error estimates ($\E(\widehat{SE})$), 95\% confidence interval coverage probability (Cov.prob), and empirical power of $\beta_{32}$ and $\beta_{33}$ are used to evaluate our proposed estimator. Let $\hat{\beta}_{3j}^{(k)}$ be the estimate of $\beta_{3j}$ from the $k$th simulated data and $\widehat{SE} (\hat{\beta}_{3j}^{(k)})$ be the corresponding standard error estimate for $j = 2, 3$ and $k = 1, \cdots, K$, the bias is approximated by $\frac{1}{K} \sum_{k=1}^{K} \hat{\beta}_{3j}^{(k)} - \beta_{3j}$, the $\mbox{SE}(\hat{\beta})$ is approximated by the sample standard deviation of $\hat{\bfbeta}_{3j}^{(k)}$ over the $K$ simulated experiments, the $\E(\widehat{SE})$ is approximated by $\frac{1}{K} \sum_{k=1}^{K} \widehat{SE} (\hat{\beta}_{3j}^{(k)})$, the 95\% confidence interval coverage probability is calculated by $\frac{1}{K} \sum_{k=1}^{K} I(\beta_{3j} \in (\hat{\beta}_{3j}^{(k)} - z_{0.025} \widehat{SE} (\hat{\beta}_{3j}^{(k)}), \hat{\beta}_{3j}^{(k)} + z_{0.025} \widehat{SE} (\hat{\beta}_{3j}^{(k)})))$, and power is calculated by $\frac{1}{K} \sum_{k=1}^{K} I([\hat{\beta}_{3j}^{(k)}/\widehat{SE} (\hat{\beta}_{3j}^{(k)})]^2 > \chi^2_1 (0.95))$, where $I(\cdot)$ is the indicator function. 

\subsection{Simulation results without model selection}
We first consider the setting in which the change point, $\tau$, is known and equal to the time of policy intervention, $t_0$.

\subsubsection{Type I error rate} \label{subsubsec:type11}
To ensure the validity of the testing procedures, we first examine the type I error rate, without model selection, for the test of level change ($H_0: \beta_{32} = 0$ vs. $H_1: \beta_{32} \ne 0$) and the test of trend change ($H_0: \beta_{33} = 0$ vs. $H_1: \beta_{33} \ne 0$) using the Wald tests introduced in Section \ref{subsec:SE}. We consider the sample sizes of $n = 60, 80, \cdots, 400$, and let $\beta_{32} = \beta_{33} = 0$.

\begin{figure}[htbp]
    \centering
    \begin{minipage}[b]{0.48\textwidth}
    \centering
    \includegraphics[width=\textwidth]{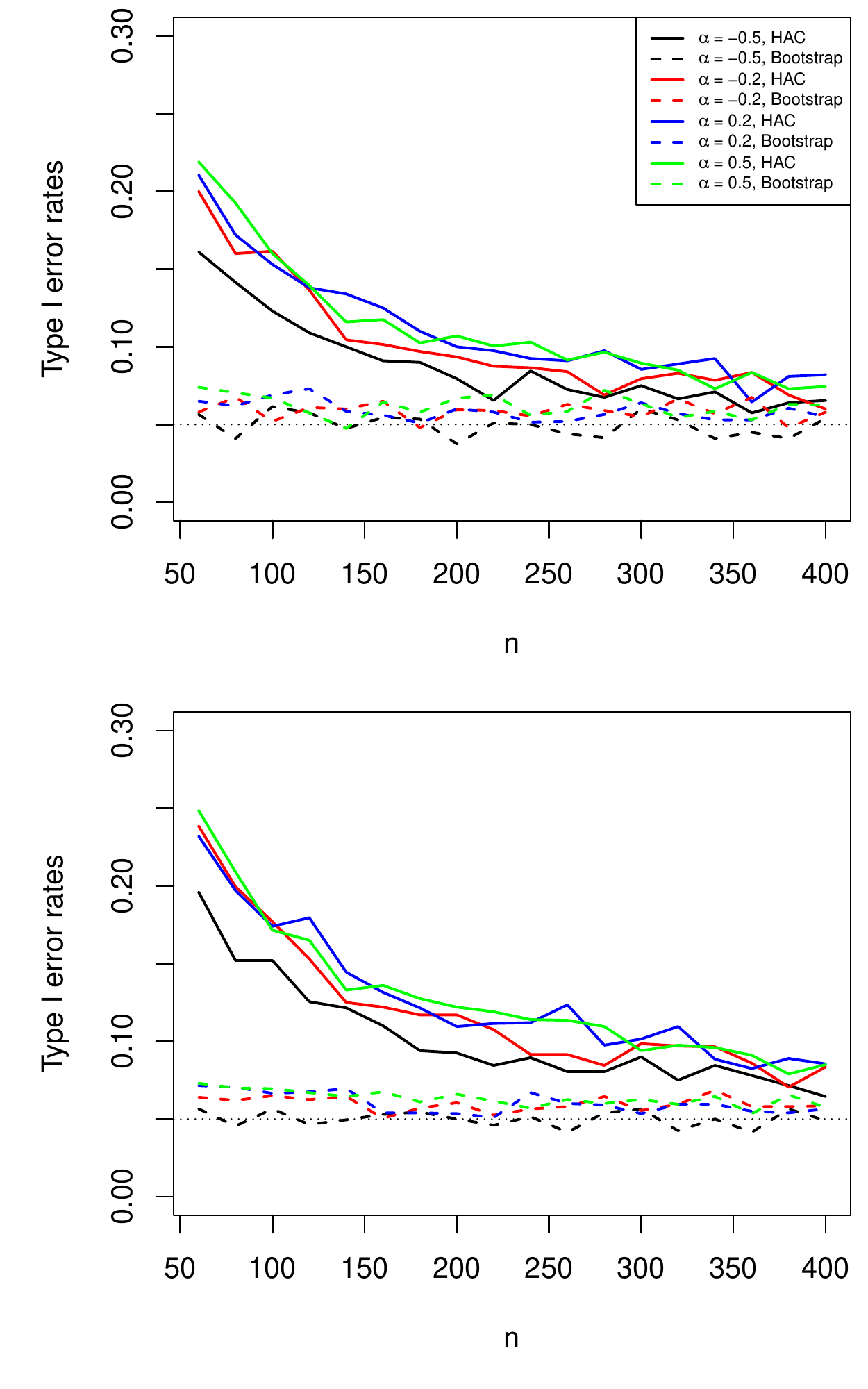}
    \caption{Empirical type I error rates using HAC and bootstrap methods (without model selection) of the test of level change (top) and the test of trend change (bottom) for data generated from zero-one inflated Beta time series with Gaussian copula and fitted by Gaussian copula models.}
    \label{typeI1}
    \end{minipage}
    \hfill
    \begin{minipage}[b]{0.48\textwidth}
    \includegraphics[width=\textwidth]{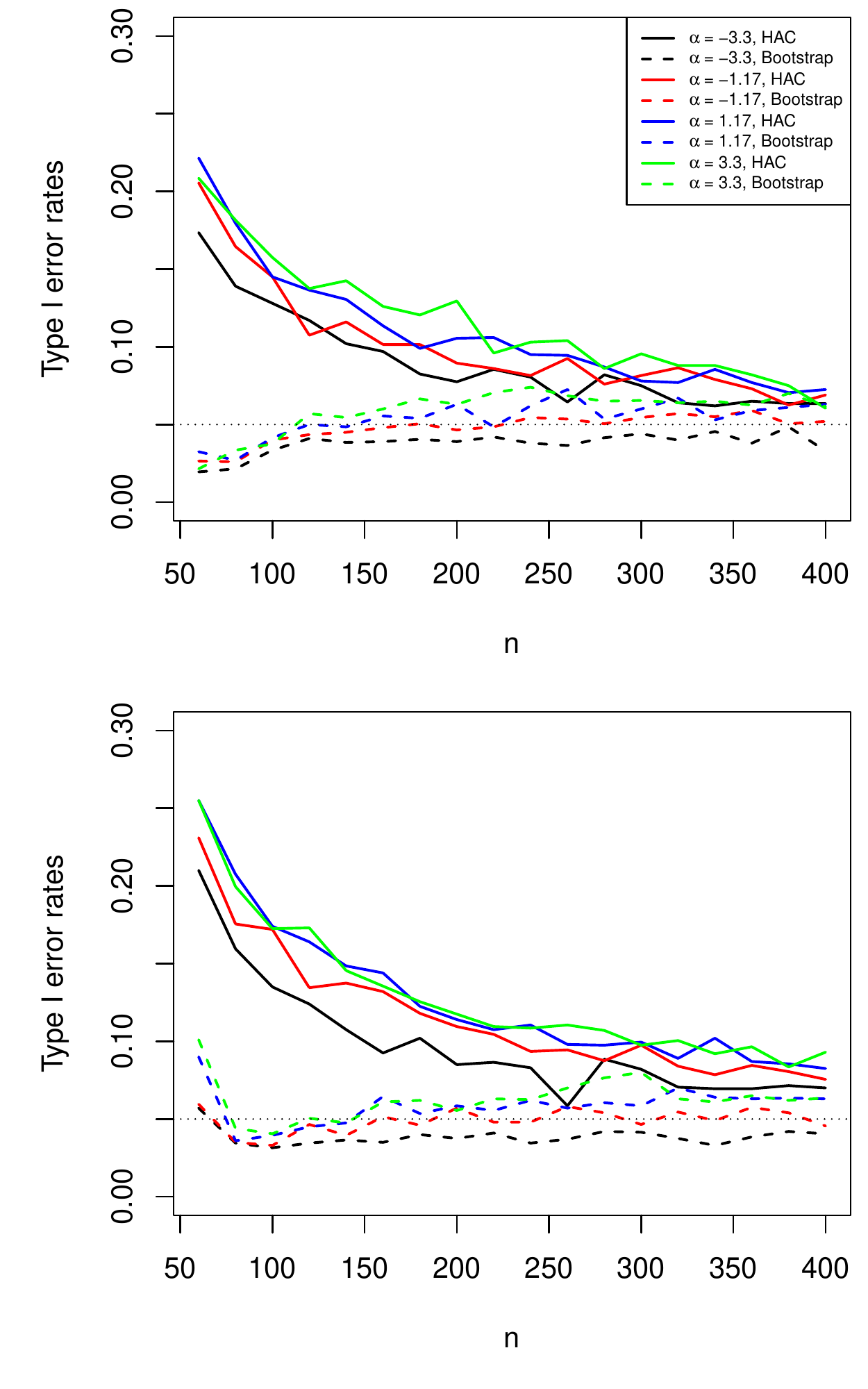}
    \caption{Empirical type I error rates using HAC and bootstrap methods (without model selection) of the test of level change (top) and the test of trend change (bottom) for data generated from zero-one inflated Beta time series with Frank copula and fitted by Gaussian copula models.}
    \label{typeI2}
    \end{minipage}
\end{figure}

Figure \ref{typeI1} illustrates the simulation results for data generated from Gaussian copula models and fitted by Gaussian copula models (correct models). For all sample sizes and copula parameter values, the empirical type I error rates for both tests (level change and trend change) using the Bootstrap methods are close to the nominal level of 0.05. However, the HAC method demonstrates a size-dependent performance, producing noticeably higher empirical Type I error rates for both tests with smaller sample sizes. When the sample size exceeds 300, the Type I error rates estimated by the HAC method improve significantly, converging towards the nominal level of 0.05. This pattern highlights the HAC method's sensitivity to sample size and its capacity for accurate estimation with sufficient data. Figure \ref{typeI2} illustrates the simulation results for data generated from Frank copula models but fitted by Gaussian copula models (misspecified models). The empirical type I error rates, except where $\rho=-1.17$ (corresponding to $\rho=-0.2$ for Gaussian copula), are in general less close to the nominal level than when models are correctly specified (Figure \ref{typeI}). Due to the model misspecification, when sample sizes are small, the empirical type I error rates of the bootstrap method are conservative for the test of level change but inflated for the test of trend change. The HAC method still shows size-dependent performance in this case, which leads to higher empirical Type I error rates for both tests when sample sizes are smaller. Comparing between the two standard error estimators, the bootstrap estimator results in closer to nominal empirical Type I error rates in all settings. 

\subsubsection{Power}
\begin{figure}[htbp]
    \centering
    \includegraphics[width=\textwidth]{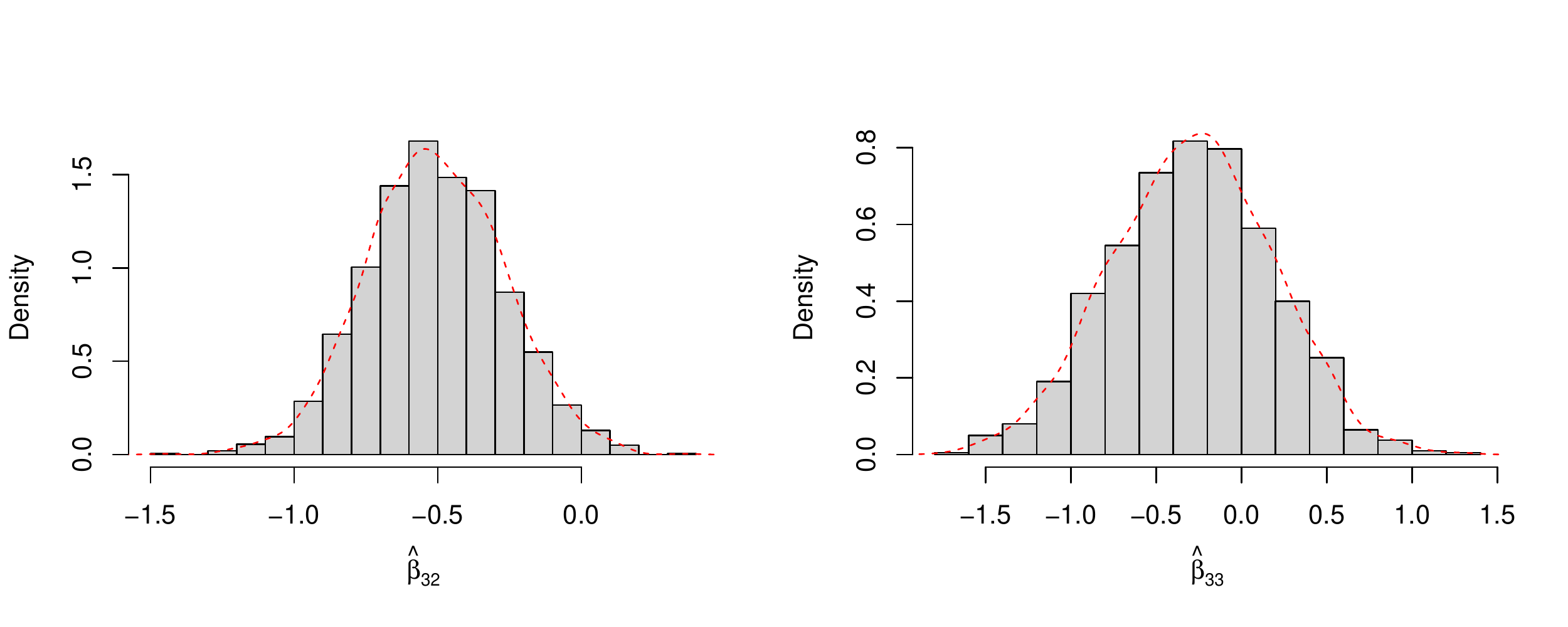}
    \caption{The empirical distributions of $\hat{\beta}_{32}$ and $\hat{\beta}_{33}$ when $n=60$}
    \label{distribution}
\end{figure}

Figure \ref{distribution} illustrates the empirical distributions of $\hat{\beta}_{32}$ and $\hat{\beta}_{33}$ when $n=60$. Both distributions are approximately normal, indicating that the proposed estimator is still valid for small sample sizes. Notice that the model misspecification on the copula model does not have an impact on the performance of point estimates on the marginal parameters because the independent likelihood estimator only depends on the marginal distribution of $\bfY$. 

We consider sample sizes of $n$ = 60, 120, or 180, and summarize the simulation results for data generated from Gaussian copula models and fitted by Gaussian copula models (correct models) in Table \ref{Result1}. For estimators ($\hat{\beta}_{32}$ and $\hat{\beta}_{33}$) with the bootstrap methods and the HAC methods, biases are close to 0 and negligible in all settings. Standard error of estimates ($\mbox{SE}(\hat{\beta})$) and means of standard error estimates ($\E(\widehat{\mbox{SE}})$) decrease, and power increases, as the sample size increases or copula parameter, $\rho,$ decreases. The variance estimator tends to underestimate its empirical counterpart (i.e., $\E(\widehat{\mbox{SE}}) < \mbox{SE}(\hat{\beta})$) when using bootstrap methods. As a result, the actual coverage probabilities of 95\% confidence intervals decrease as the sample size increases with this method. For estimates with HAC methods, the coverage probabilities increase as the sample size increases, but they are always lower than the coverage probabilities with bootstrap methods. Due to the relatively small magnitude of $\beta_{33}$ (compared with $\beta_{32}$), the power of $\beta_{33}$ is much smaller than the power of $\beta_{32}$ in all settings.

Results for the misspecified models (data generated from Frank copula models but fitted by Gaussian copula models) are summarized in Table \ref{Result2}. Overall, our proposed estimator is robust against model misspecification. Similar to the correctly specified models, both estimators are almost unbiased. Moreover, $\mbox{SE}(\hat{\beta})$ and $\E(\widehat{\mbox{SE}})$ decrease, and power increases, as sample size increases or $\rho$ decreases. When the sample size is small ($n = 60$), the variance estimator of $\beta_{33}$ with HAC method overestimates its empirical counterpart (i.e., $\E(\widehat{\mbox{SE}}) > \mbox{SE}(\hat{\beta})$), leading to low power in these scenarios. Although the power utilized by the HAC method is consistently lower than that utilized by the bootstrap method, it is likely due to the inflated empirical type I error rates of the HAC method as illustrated in the previous section. 

\subsection{Simulation results with model selection}
We then consider the setting in which the change point is assumed unknown but still set to the time of policy intervention, and the model selection method described in Section \ref{subsec:selection} is used for change point selection. For illustrative purposes, we only consider five candidate values of the change point in the model selection: $\tau-2$, $\tau-1$, $\tau$, $\tau+1$, and $\tau+2$. In practice, we may want to include as many candidate values as possible to ensure the true value is included in the candidate set.

\subsubsection{Type I error rate}
We first investigate the type I error rate after model selection. Similar to Section \ref{subsubsec:type11}, we also consider the sample sizes of $n = 60, 80, \cdots, 400$. 

\begin{figure}[htbp]
    \centering
    \begin{minipage}[b]{0.48\textwidth}
    \centering
    \includegraphics[width=\textwidth]{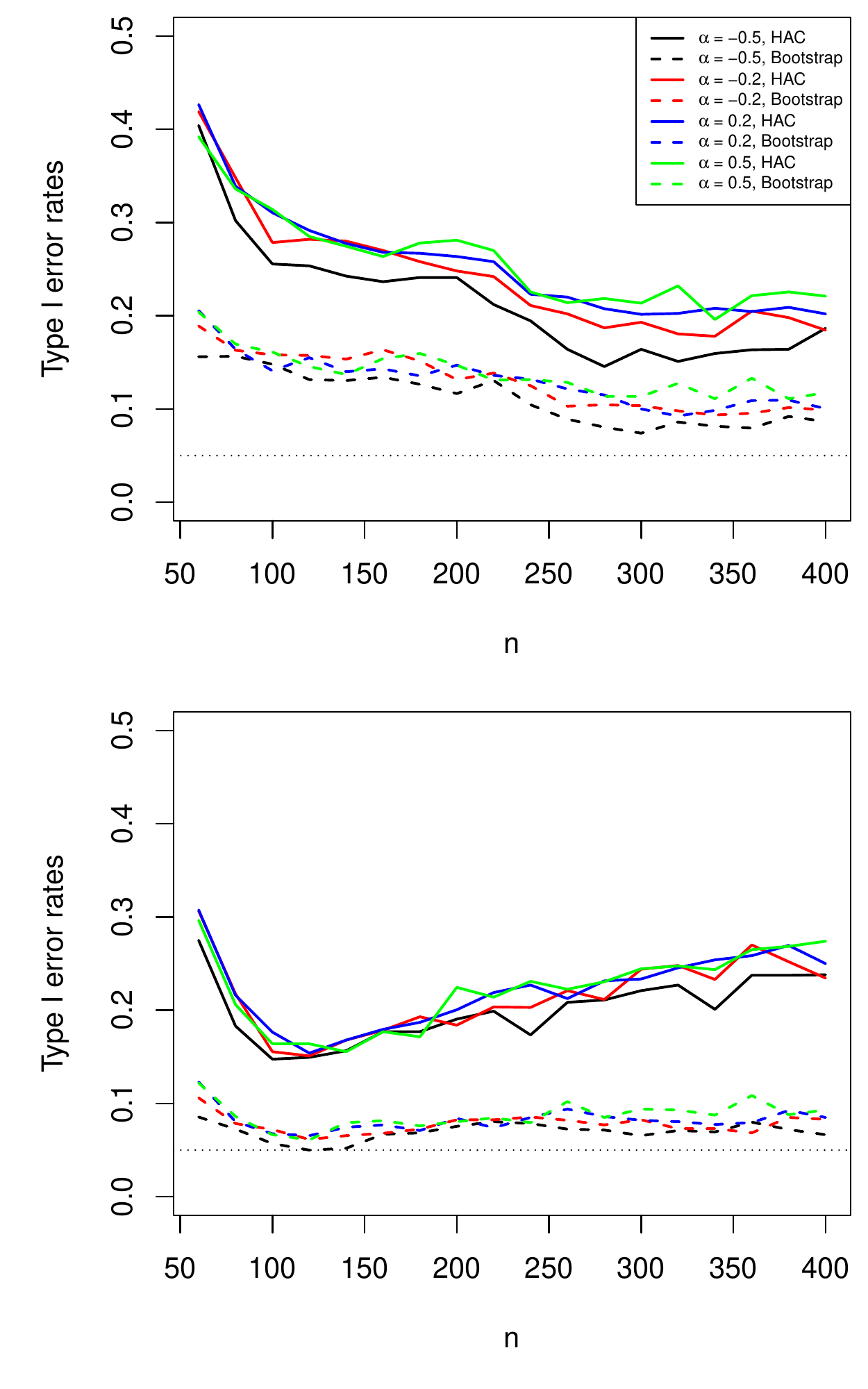}
    \caption{Empirical type I error rates using HAC and bootstrap methods (with model selection) of the test of level change (top) and the test of trend change (bottom) for data generated from zero-one inflated Beta time series with Gaussian copula and fitted by Gaussian copula models.}
    \label{typeI3}
    \end{minipage}
    \hfill
    \begin{minipage}[b]{0.48\textwidth}
    \includegraphics[width=\textwidth]{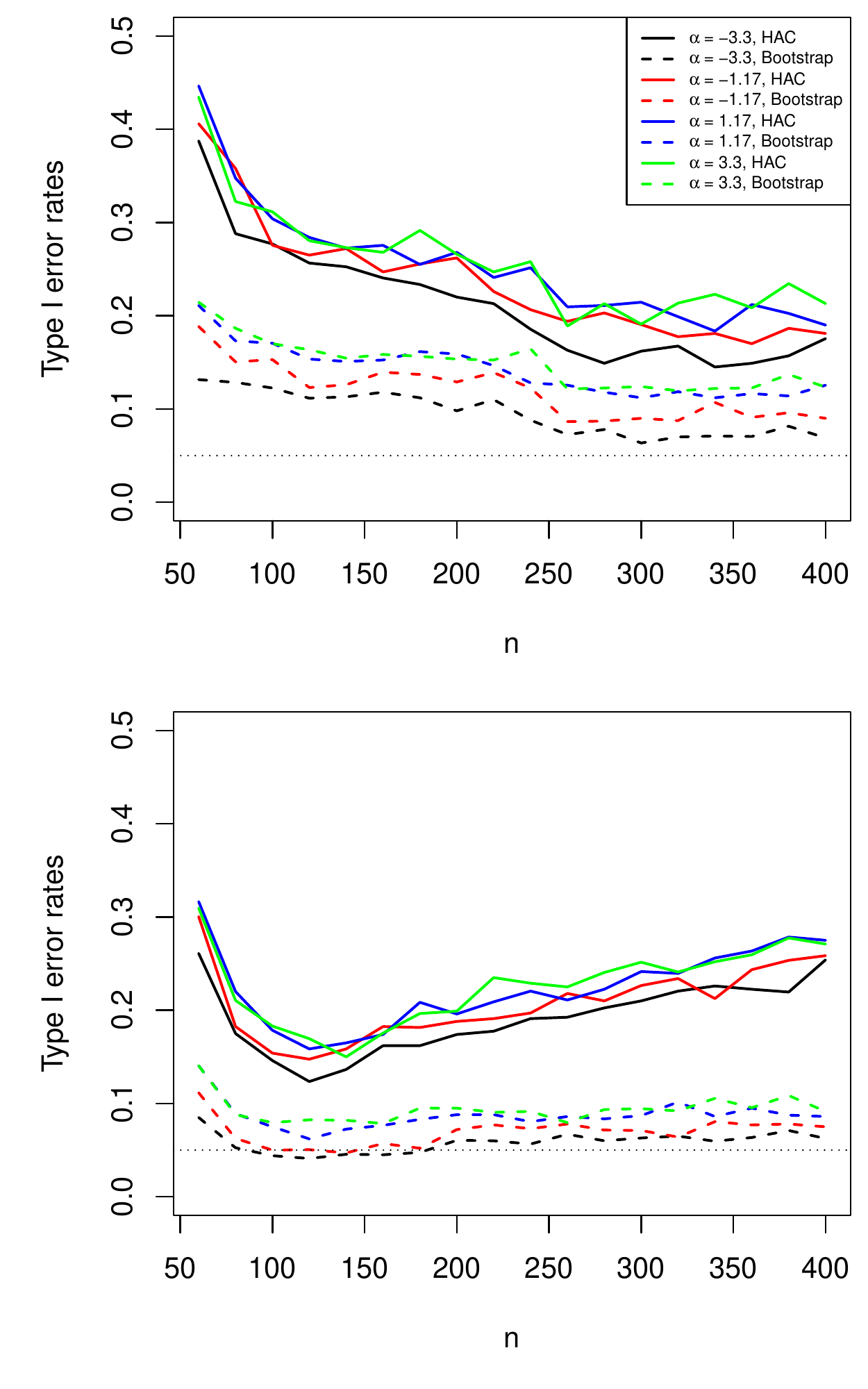}
    \caption{Empirical type I error rates using HAC and bootstrap methods (with model selection) of the test of level change (top) and the test of trend change (bottom) for data generated from zero-one inflated Beta time series with Frank copula and fitted by Gaussian copula models.}
    \label{typeI4}
    \end{minipage}
\end{figure}

Figure \ref{typeI3} illustrates the simulation results for data generated from Gaussian copula models and fitted by Gaussian copula models (correct models), and Figure \ref{typeI4} illustrates the simulation results for data generated from Frank copula models but fitted by Gaussian copula models (misspecified models). When using bootstrap methods, the empirical type I error rates are conservative for the test of level change but inflated for the test of trend change when sample sizes are small. In general, the empirical type I error rates with bootstrap methods after model selection are more inflated than their counterparts without model selection, because the selection method tends to select a model that maximizes the estimated effects of level and trend changes. In the case of HAC methods, when the sample size exceeds 150, the empirical type I error rates tend to increase with the sample size for the test of level change.

\subsubsection{Power}
Table \ref{Result3} presents the simulation results for data generated from Gaussian copula and fitted by Gaussian copula models (correct models) with sample sizes $n$ = 60, 120, or 180 are considered. Contrary to the results without model selection (Table \ref{Result1}), the proposed estimator is biased, and the bias decreases as the sample size increases. As a result, the coverage probabilities of 95\% CIs are lower than the corresponding coverage probabilities without model selection. Similar trends are observed for other measures:  $\mbox{SE}(\hat{\beta})$ and $\E(\widehat{\mbox{SE}})$ decrease as sample size increases or copula parameter $\rho$ decreases, and coverage probabilities and power increase as sample size increases.

Simulation results for data generated from Frank copula and fitted by Gaussian copula models (misspecified models) are summarized in Table \ref{Result4}. Biases of both estimators are similar to the results of the correct models with model selection (Table \ref{Result3}), which again verifies the robustness of the Gaussian copula model. $\mbox{SE}(\hat{\beta})$ and $\E(\widehat{\mbox{SE}})$ decrease, and power increases, as sample size increases or copula parameter $\rho$ decreases.

\section{ITS analysis of the intervention effect on patient `pain management' scores} \label{sec:app}
Patient `pain management' scores of a single hospital are modeled using our proposed MZOIBTS model. Figure \ref{Data} illustrates the data used for this analysis.

We consider the models suggested in Section \ref{subsec:ITS}, where the marginal mean of the time series (which is the main model in ITS analysis) is assumed as in model (\ref{segment}) with $T(t) = t$. We present results based on the Gaussian copula, but other copula functions result in similar conclusions. We only consider the bootstrap standard error estimator because our simulation results suggest that the Type I error rates of the HAC estimator are inflated for small sample sizes. We assume the change point of the intervention effect can happen anytime between January 2010 and January 2011, i.e., 6 months before and after the formal intervention time. Even though we observed inflated type I error rates in our simulation studies when the change point is being estimated, in our data setting, we have to do it due to the nature of the intervention.

Figure \ref{Data} shows the fitted mean function, and the resulting parameter estimates are given in Table \ref{App Results}. The estimated change point occurred in October 2010, about 4 months after the formal implementation time, indicating that the implementation of the intervention requires a period of time to reach its full extent. The estimated `pain management' score is initiated at 62.9\% and gradually increases to 75.4\% before the estimated change point (October 2010). A 6\% increase in the estimated `pain management' score is observed immediately after the estimated change point, but the score then gradually decreases to 76.8\% by the end of the follow-up period. However, the two key parameters measuring level change ($\beta_{32}$) and trend change ($\beta_{33}$) are not statistically significant at the $\alpha = 0.05$ level ($p$-values: 0.265 and 0.103 respectively). 

Our results reveal a significant increase in the dispersion parameter $\phi_t$ ($\beta_{41} = 0.795$, $p$-value = 0.024). According to the second equation in (\ref{moment}), an increase in $\phi_t$ results in a reduction in the standard deviation of the outcome. In this study, the average estimated standard deviation of $Y_t$ decreases from 0.143 before intervention to 0.110 after the change point.

\section{Discussion} \label{sec:con}
In this article, motivated by ITS analysis, we propose a copula-based time series model for zero-one-inflated proportional data analysis and develop a two-stage estimation procedure, where regression parameters are estimated by maximizing the composite log-likelihood function and two estimators, HAC and parametric bootstrap, are proposed for standard error estimation. Our simulation results reveal that the bootstrap estimator is valid for small sample sizes. We apply the proposed MZOIBTS model to study the impact of a new nursing care delivery mode on the patients `pain management' scores in a single hospital unit. Although we do not find any significant changes in the level and trend of `pain management' scores after the intervention, the variance of the scores significantly decreases after the change point. As pointed out by \cite{cruz2017robust}, the reduction in the variance can be considered a positive result of the intervention since it leads to more stable scores.

We focus on the estimation and inference of the marginal model parameters, while the copula parameter is considered as nuisance and is being estimated separately. Completely nonparametric standard error estimator ignoring the copula information, such as HAC, suffers from a slow rate of convergence, which results in inflated Type I error rates. Thus, we suggest using the parametric bootstrap method for standard error estimation when the sample size is small. Although we can select the optimal copula function by maximizing the second step objective function, it can be computationally expensive due to the slow convergence when estimating some of the copula functions such as Clayton. We suggest considering the Gaussian copula as the default choice. Our simulation results have verified the robustness of the use of the Gaussian copula function against model misspecification. Moreover, different copula functions also result in similar inferences and conclusions in our real data analysis.

Our analysis does not detect any level or trend changes after intervention. This can be due to the ceiling effect, where the mean score will stop increasing after it reaches its maximum. On the one hand, our proposed model partially addresses the ceiling effect by bounding the marginal means of outcomes to 0 and 1. On the other hand, however, we assume a nonstationary time series with the deterministic specified as the generalized segmented linear regression model (\ref{segment}). It assumes the logit-transformed marginal mean $v_t$ keeps increasing/decreasing as $t$ increases if $\beta_{31} + \beta_{33} \ne 0$, but in practice, it is possible that the time series will become stationary after $v_t$ reaches a certain level. Thus, it is of great interest to investigate new models to include this ceiling/flooring effect.  

Finally, some other models can also be considered for such bounded outcomes. For example, the tilted Beta distribution proposed by \cite{hahn2021regression} is an alternative distribution for bounded outcomes with multiple zeros or ones. We can also assume the outcomes are censored at zero and one, and marginalized Tobit regression models \cite{wang2017natural} can be used to analyze such data. It is of interest to incorporate serial dependence in these models.

\appendix
\section{Composite score function}
In this Appendix, we give the composite score function
$u_t(\bftheta)$, which can be expressed as
\be \label{score}
u_t(\bftheta) = \left[ \frac{\partial l_t(\bftheta; y_t)}{\partial \bfbeta_1}, \frac{\partial l_t(\bftheta; y_t)}{\partial \bfbeta_2}, \frac{\partial l_t(\bftheta; y_t)}{\partial \bfbeta_3}, \frac{\partial l_t(\bftheta; y_t)}{\partial \bfbeta_4} \right]^\top,
\ee
where
\ba
\frac{\partial l_t(\bftheta; y_t)}{\partial \bfbeta_1} &=& -\frac{\exp(\bfx_{1t}^\top \bfbeta_1)}{1 + \exp(\bfx_{1t}^\top \bfbeta_1)} \bfx_{1t}^\top + I(y_t>0) \bfx_{1t}^\top \\ &~& + I(0<y_t<1) \left\{ h'_{t, \bfbeta_1} + \mu'_{t, \bfbeta_1} \exp(\bfx_{4t}^\top \bfbeta_4) \text{logit}(y_t) \bfx_{1t}^\top \right\}, \\
\frac{\partial l_t(\bftheta; y_t)}{\partial \bfbeta_2} &=& -I(y_t>0) \frac{\exp(\bfx_{2t}^\top \bfbeta_2)}{1 + \exp(\bfx_{2t}^\top \bfbeta_2)} \bfx_{2t}^\top + I(y_t=1) \bfx_{2t}^\top \\ &~& + I(0<y_t<1) \left\{ h'_{t, \bfbeta_2} + \mu'_{t, \bfbeta_2} \exp(\bfx_{4t}^\top \bfbeta_4) \text{logit}(y_t) \bfx_{2t}^\top \right\}, \\ 
\frac{\partial l_t(\bftheta; y_t)}{\partial \bfbeta_3} &=& I(0<y_t<1) \left\{ h'_{t, \bfbeta_3} + \mu'_{t, \bfbeta_3} \exp(\bfx_{4t}^\top \bfbeta_4) \text{logit}(y_t) \bfx_{3t}^\top \right\}, \\
\frac{\partial l_t(\bftheta; y_t)}{\partial \bfbeta_4} &=& I(0<y_t<1) \left\{ h'_{t, \bfbeta_4} + \mu_t \exp(\bfx_{4t}^\top \bfbeta_4) \text{logit}(y_t) \bfx_{4t}^\top + \exp(\bfx_{4t}^\top \bfbeta_4) \log(1-y_t) \bfx_{4t}^\top \right\},
\ea
and
\ba
\mu'_{t, \bfbeta_1} &=& -\frac{(1 + \exp(\bfx_{2t}^\top \bfbeta_2)) \exp(-\bfx_{1t}^\top \bfbeta_1)}{1 + \exp(-\bfx_{3t}^\top \bfbeta_3)}, \\
\mu'_{t, \bfbeta_2} &=& \frac{\exp(\bfx_{2t}^\top \bfbeta_2) (1 + \exp(-\bfx_{1t}^\top \bfbeta_1))}{1 + \exp(-\bfx_{3t}^\top \bfbeta_3)} - \exp(\bfx_{2t}^\top \bfbeta_2), \\
\mu'_{t, \bfbeta_3} &=& -\frac{(1 + \exp(\bfx_{2t}^\top \bfbeta_2)) (1 + \exp(-\bfx_{1t}^\top \bfbeta_1)) \exp(-\bfx_{3t}^\top \bfbeta_3)}{(1 + \exp(-\bfx_{3t}^\top \bfbeta_3))^2},
\ea
and
\ba
h'_{t, \bfbeta_1} &=& -\left[ \psi(\mu_t \exp(\bfx_{4t}^\top \bfbeta_4)) -\psi((1 - \mu_t) \exp(\bfx_{4t}^\top \bfbeta_4)) \right] \exp(\bfx_{4t}^\top \bfbeta_4) \mu'_{t, \bfbeta_1} \bfx_{1t}^\top, \\
h'_{t, \bfbeta_2} &=& -\left[ \psi(\mu_t \exp(\bfx_{4t}^\top \bfbeta_4)) -\psi((1 - \mu_t) \exp(\bfx_{4t}^\top \bfbeta_4)) \right] \exp(\bfx_{4t}^\top \bfbeta_4) \mu'_{t, \bfbeta_2} \bfx_{2t}^\top, \\
h'_{t, \bfbeta_3} &=& -\left[ \psi(\mu_t \exp(\bfx_{4t}^\top \bfbeta_4)) -\psi((1 - \mu_t) \exp(\bfx_{4t}^\top \bfbeta_4)) \right] \exp(\bfx_{4t}^\top \bfbeta_4) \mu'_{t, \bfbeta_3} \bfx_{3t}^\top, \\
h'_{t, \bfbeta_4} &=& \left[ \psi(\exp(\bfx_{4t}^\top \bfbeta_4)) - \psi(\mu_t \exp(\bfx_{4t}^\top \bfbeta_4)) \mu_t - \psi((1 - \mu_t) \exp(\bfx_{4t}^\top \bfbeta_4)) (1 - \mu_t) \right] \exp(\bfx_{4t}^\top \bfbeta_4) \bfx_{4t}^\top,
\ea
where $\psi(a) = \frac{\partial}{\partial a} \log \Gamma(a)$ is the digamma function of $a$. 

\section*{Supplementary Materials}
The R code to implement the proposed algorithm is provided at \url{https://github.com/shyye008/MZOIBTS}.

\section*{Acknowledgements}
This work was supported by the Knight Cancer Institute Biostatistics Shared Resource at Oregon Health and Science University (NCI Cancer Center Support Grant P30 CA069533). 

\bibliographystyle{elsarticle-harv}
\biboptions{authoryear}
\bibliography{ref}

\begin{thebibliography}{44}
\expandafter\ifx\csname natexlab\endcsname\relax\def\natexlab#1{#1}\fi
\providecommand{\url}[1]{\texttt{#1}}
\providecommand{\href}[2]{#2}
\providecommand{\path}[1]{#1}
\providecommand{\DOIprefix}{doi:}
\providecommand{\ArXivprefix}{arXiv:}
\providecommand{\URLprefix}{URL: }
\providecommand{\Pubmedprefix}{pmid:}
\providecommand{\doi}[1]{\href{http://dx.doi.org/#1}{\path{#1}}}
\providecommand{\Pubmed}[1]{\href{pmid:#1}{\path{#1}}}
\providecommand{\bibinfo}[2]{#2}
\ifx\xfnm\relax \def\xfnm[#1]{\unskip,\space#1}\fi
%Type = Article
\bibitem[{Abdel-Karim(2017)}]{abdel2017extended}
\bibinfo{author}{Abdel-Karim, A.H.}, \bibinfo{year}{2017}.
\newblock \bibinfo{title}{Extended zero-one inflated beta and adjusted
  three-part regression models for proportional data analysis}.
\newblock \bibinfo{journal}{Communications in Statistics-Simulation and
  Computation} \bibinfo{volume}{46}, \bibinfo{pages}{6155--6172}.
%Type = Article
\bibitem[{Alqawba and Diawara(2021)}]{alqawba2021copula}
\bibinfo{author}{Alqawba, M.}, \bibinfo{author}{Diawara, N.},
  \bibinfo{year}{2021}.
\newblock \bibinfo{title}{Copula-based markov zero-inflated count time series
  models with application}.
\newblock \bibinfo{journal}{Journal of Applied Statistics}
  \bibinfo{volume}{48}, \bibinfo{pages}{786--803}.
%Type = Article
\bibitem[{Bender et~al.(2012)Bender, Connelly, Glaser and
  Brown}]{Bender:2012ch}
\bibinfo{author}{Bender, M.}, \bibinfo{author}{Connelly, C.D.},
  \bibinfo{author}{Glaser, D.}, \bibinfo{author}{Brown, C.},
  \bibinfo{year}{2012}.
\newblock \bibinfo{title}{Clinical nurse leader impact on microsystem care
  quality}.
\newblock \bibinfo{journal}{Nursing Research} \bibinfo{volume}{61},
  \bibinfo{pages}{326--332}.
%Type = Inproceedings
\bibitem[{Bender et~al.(2015)Bender, Murphy, Thomas, Kaminski and
  Smith}]{bender2015clinical}
\bibinfo{author}{Bender, M.}, \bibinfo{author}{Murphy, E.},
  \bibinfo{author}{Thomas, T.}, \bibinfo{author}{Kaminski, J.},
  \bibinfo{author}{Smith, B.}, \bibinfo{year}{2015}.
\newblock \bibinfo{title}{Clinical nurse leader integration into care delivery
  microsystems: quality and safety outcomes at the unit and organization
  level}, in: \bibinfo{booktitle}{Academy Health Annual Research Meeting}.
%Type = Article
\bibitem[{Bender et~al.(2019)Bender, Murphy, Cruz and Ombao}]{bender2019system}
\bibinfo{author}{Bender, M.}, \bibinfo{author}{Murphy, E.A.},
  \bibinfo{author}{Cruz, M.}, \bibinfo{author}{Ombao, H.},
  \bibinfo{year}{2019}.
\newblock \bibinfo{title}{System-and unit-level care quality outcome
  improvements after integrating clinical nurse leaders into frontline care
  delivery}.
\newblock \bibinfo{journal}{JONA: the Journal of Nursing Administration}
  \bibinfo{volume}{49}, \bibinfo{pages}{315--322}.
%Type = Article
\bibitem[{Bender et~al.(2017)Bender, Williams, Su and Hites}]{Bender2017}
\bibinfo{author}{Bender, M.}, \bibinfo{author}{Williams, M.},
  \bibinfo{author}{Su, W.}, \bibinfo{author}{Hites, L.}, \bibinfo{year}{2017}.
\newblock \bibinfo{title}{Refining and validating a conceptual model of
  clinical nurse leader integrated care delivery}.
\newblock \bibinfo{journal}{Journal of Advanced Nursing} \bibinfo{volume}{73},
  \bibinfo{pages}{448--464}.
%Type = Article
\bibitem[{Bernal et~al.(2017)Bernal, Cummins and
  Gasparrini}]{bernal2017interrupted}
\bibinfo{author}{Bernal, J.L.}, \bibinfo{author}{Cummins, S.},
  \bibinfo{author}{Gasparrini, A.}, \bibinfo{year}{2017}.
\newblock \bibinfo{title}{Interrupted time series regression for the evaluation
  of public health interventions: a tutorial}.
\newblock \bibinfo{journal}{International journal of epidemiology}
  \bibinfo{volume}{46}, \bibinfo{pages}{348--355}.
%Type = Article
\bibitem[{Casarin et~al.(2012)Casarin, Dalla~Valle and
  Leisen}]{casarin2012bayesian}
\bibinfo{author}{Casarin, R.}, \bibinfo{author}{Dalla~Valle, L.},
  \bibinfo{author}{Leisen, F.}, \bibinfo{year}{2012}.
\newblock \bibinfo{title}{Bayesian model selection for beta autoregressive
  processes}.
\newblock \bibinfo{journal}{Bayesian Analysis} \bibinfo{volume}{7},
  \bibinfo{pages}{385--410}.
%Type = Article
\bibitem[{Chai et~al.(2018)Chai, Jiang, Lin and Liu}]{chai2018marginalized}
\bibinfo{author}{Chai, H.}, \bibinfo{author}{Jiang, H.}, \bibinfo{author}{Lin,
  L.}, \bibinfo{author}{Liu, L.}, \bibinfo{year}{2018}.
\newblock \bibinfo{title}{A marginalized two-part beta regression model for
  microbiome compositional data}.
\newblock \bibinfo{journal}{PLoS computational biology} \bibinfo{volume}{14},
  \bibinfo{pages}{e1006329}.
%Type = Article
\bibitem[{Chandler and Bate(2007)}]{chandler2007inference}
\bibinfo{author}{Chandler, R.E.}, \bibinfo{author}{Bate, S.},
  \bibinfo{year}{2007}.
\newblock \bibinfo{title}{Inference for clustered data using the independence
  loglikelihood}.
\newblock \bibinfo{journal}{Biometrika} \bibinfo{volume}{94},
  \bibinfo{pages}{167--183}.
%Type = Book
\bibitem[{Cook et~al.(2002)Cook, Campbell and Shadish}]{cook2002experimental}
\bibinfo{author}{Cook, T.D.}, \bibinfo{author}{Campbell, D.T.},
  \bibinfo{author}{Shadish, W.}, \bibinfo{year}{2002}.
\newblock \bibinfo{title}{Experimental and quasi-experimental designs for
  generalized causal inference}.
\newblock \bibinfo{publisher}{Houghton Mifflin Boston, MA}.
%Type = Article
\bibitem[{Cruz et~al.(2017)Cruz, Bender and Ombao}]{cruz2017robust}
\bibinfo{author}{Cruz, M.}, \bibinfo{author}{Bender, M.},
  \bibinfo{author}{Ombao, H.}, \bibinfo{year}{2017}.
\newblock \bibinfo{title}{A robust interrupted time series model for analyzing
  complex health care intervention data}.
\newblock \bibinfo{journal}{Statistics in medicine} \bibinfo{volume}{36},
  \bibinfo{pages}{4660--4676}.
%Type = Article
\bibitem[{Cruz et~al.(2019)Cruz, Gillen, Bender and Ombao}]{cruz2019assessing}
\bibinfo{author}{Cruz, M.}, \bibinfo{author}{Gillen, D.L.},
  \bibinfo{author}{Bender, M.}, \bibinfo{author}{Ombao, H.},
  \bibinfo{year}{2019}.
\newblock \bibinfo{title}{Assessing health care interventions via an
  interrupted time series model: study power and design considerations}.
\newblock \bibinfo{journal}{Statistics in medicine} \bibinfo{volume}{38},
  \bibinfo{pages}{1734--1752}.
%Type = Article
\bibitem[{Da-Silva and Migon(2016)}]{da2016hierarchical}
\bibinfo{author}{Da-Silva, C.Q.}, \bibinfo{author}{Migon, H.S.},
  \bibinfo{year}{2016}.
\newblock \bibinfo{title}{Hierarchical dynamic beta model}.
\newblock \bibinfo{journal}{Revstat Statistical Journal} \bibinfo{volume}{14},
  \bibinfo{pages}{49--73}.
%Type = Article
\bibitem[{Davis et~al.(2003)Davis, Dunsmuir and Streett}]{davis2003observation}
\bibinfo{author}{Davis, R.A.}, \bibinfo{author}{Dunsmuir, W.T.},
  \bibinfo{author}{Streett, S.B.}, \bibinfo{year}{2003}.
\newblock \bibinfo{title}{Observation-driven models for poisson counts}.
\newblock \bibinfo{journal}{Biometrika} \bibinfo{volume}{90},
  \bibinfo{pages}{777--790}.
%Type = Book
\bibitem[{Davison and Hinkley(1997)}]{davison1997bootstrap}
\bibinfo{author}{Davison, A.C.}, \bibinfo{author}{Hinkley, D.V.},
  \bibinfo{year}{1997}.
\newblock \bibinfo{title}{Bootstrap methods and their application}.
\newblock \bibinfo{number}{1}, \bibinfo{publisher}{Cambridge university press}.
%Type = Article
\bibitem[{van Doormaal et~al.(2009)van Doormaal, van~den Bemt, Zaal, Egberts,
  Lenderink, Kosterink, Haaijer-Ruskamp and Mol}]{van2009influence}
\bibinfo{author}{van Doormaal, J.E.}, \bibinfo{author}{van~den Bemt, P.M.},
  \bibinfo{author}{Zaal, R.J.}, \bibinfo{author}{Egberts, A.C.},
  \bibinfo{author}{Lenderink, B.W.}, \bibinfo{author}{Kosterink, J.G.},
  \bibinfo{author}{Haaijer-Ruskamp, F.M.}, \bibinfo{author}{Mol, P.G.},
  \bibinfo{year}{2009}.
\newblock \bibinfo{title}{The influence that electronic prescribing has on
  medication errors and preventable adverse drug events: an interrupted
  time-series study}.
\newblock \bibinfo{journal}{Journal of the American Medical Informatics
  Association} \bibinfo{volume}{16}, \bibinfo{pages}{816--825}.
%Type = Article
\bibitem[{Fang and Ma(2013)}]{fang2013three}
\bibinfo{author}{Fang, K.}, \bibinfo{author}{Ma, S.}, \bibinfo{year}{2013}.
\newblock \bibinfo{title}{Three-part model for fractional response variables
  with application to chinese household health insurance coverage}.
\newblock \bibinfo{journal}{Journal of Applied Statistics}
  \bibinfo{volume}{40}, \bibinfo{pages}{925--940}.
%Type = Article
\bibitem[{Ferrari and Cribari-Neto(2004)}]{ferrari2004beta}
\bibinfo{author}{Ferrari, S.}, \bibinfo{author}{Cribari-Neto, F.},
  \bibinfo{year}{2004}.
\newblock \bibinfo{title}{Beta regression for modelling rates and proportions}.
\newblock \bibinfo{journal}{Journal of applied statistics}
  \bibinfo{volume}{31}, \bibinfo{pages}{799--815}.
%Type = Article
\bibitem[{Fokianos et~al.(2009)Fokianos, Rahbek and
  Tj{\o}stheim}]{fokianos2009poisson}
\bibinfo{author}{Fokianos, K.}, \bibinfo{author}{Rahbek, A.},
  \bibinfo{author}{Tj{\o}stheim, D.}, \bibinfo{year}{2009}.
\newblock \bibinfo{title}{Poisson autoregression}.
\newblock \bibinfo{journal}{Journal of the American Statistical Association}
  \bibinfo{volume}{104}, \bibinfo{pages}{1430--1439}.
%Type = Article
\bibitem[{Gao and Song(2010)}]{gao2010composite}
\bibinfo{author}{Gao, X.}, \bibinfo{author}{Song, P.X.K.},
  \bibinfo{year}{2010}.
\newblock \bibinfo{title}{Composite likelihood bayesian information criteria
  for model selection in high-dimensional data}.
\newblock \bibinfo{journal}{Journal of the American Statistical Association}
  \bibinfo{volume}{105}, \bibinfo{pages}{1531--1540}.
%Type = Article
\bibitem[{Godambe(1960)}]{godambe1960optimum}
\bibinfo{author}{Godambe, V.P.}, \bibinfo{year}{1960}.
\newblock \bibinfo{title}{An optimum property of regular maximum likelihood
  estimation}.
\newblock \bibinfo{journal}{The Annals of Mathematical Statistics}
  \bibinfo{volume}{31}, \bibinfo{pages}{1208--1211}.
%Type = Article
\bibitem[{Gong and Samaniego(1981)}]{gong1981pseudo}
\bibinfo{author}{Gong, G.}, \bibinfo{author}{Samaniego, F.J.},
  \bibinfo{year}{1981}.
\newblock \bibinfo{title}{Pseudo maximum likelihood estimation: theory and
  applications}.
\newblock \bibinfo{journal}{The Annals of Statistics} ,
  \bibinfo{pages}{861--869}.
%Type = Article
\bibitem[{Guolo and Varin(2014)}]{guolo2014beta}
\bibinfo{author}{Guolo, A.}, \bibinfo{author}{Varin, C.}, \bibinfo{year}{2014}.
\newblock \bibinfo{title}{Beta regression for time series analysis of bounded
  data, with application to canada google{\textregistered} flu trends}.
\newblock \bibinfo{journal}{The Annals of Applied Statistics}
  \bibinfo{volume}{8}, \bibinfo{pages}{74--88}.
%Type = Article
\bibitem[{Hahn(2021)}]{hahn2021regression}
\bibinfo{author}{Hahn, E.D.}, \bibinfo{year}{2021}.
\newblock \bibinfo{title}{Regression modelling with the tilted beta
  distribution: A bayesian approach}.
\newblock \bibinfo{journal}{Canadian Journal of Statistics}
  \bibinfo{volume}{49}, \bibinfo{pages}{262--282}.
%Type = Book
\bibitem[{Joe(1997)}]{joe1997multivariate}
\bibinfo{author}{Joe, H.}, \bibinfo{year}{1997}.
\newblock \bibinfo{title}{Multivariate models and multivariate dependence
  concepts}.
\newblock \bibinfo{publisher}{CRC press}.
%Type = Article
\bibitem[{Joe(2005)}]{joe2005asymptotic}
\bibinfo{author}{Joe, H.}, \bibinfo{year}{2005}.
\newblock \bibinfo{title}{Asymptotic efficiency of the two-stage estimation
  method for copula-based models}.
\newblock \bibinfo{journal}{Journal of multivariate Analysis}
  \bibinfo{volume}{94}, \bibinfo{pages}{401--419}.
%Type = Book
\bibitem[{Joe(2014)}]{joe2014dependence}
\bibinfo{author}{Joe, H.}, \bibinfo{year}{2014}.
\newblock \bibinfo{title}{Dependence modeling with copulas}.
\newblock \bibinfo{publisher}{CRC press}.
%Type = Techreport
\bibitem[{Joe and Xu(1996)}]{joe1996estimation}
\bibinfo{author}{Joe, H.}, \bibinfo{author}{Xu, J.J.}, \bibinfo{year}{1996}.
\newblock \bibinfo{title}{The estimation method of inference functions for
  margins for multivariate models}.
\newblock \bibinfo{type}{Technical Report} \bibinfo{number}{166}. Department of
  Statistics, University of British Columbia.
%Type = Article
\bibitem[{Kavanagh et~al.(2012)Kavanagh, Cimiotti, Abusalem and
  Coty}]{kavanagh}
\bibinfo{author}{Kavanagh, K.T.}, \bibinfo{author}{Cimiotti, J.P.},
  \bibinfo{author}{Abusalem, S.}, \bibinfo{author}{Coty, M.B.},
  \bibinfo{year}{2012}.
\newblock \bibinfo{title}{Moving healthcare quality forward with
  nursing-sensitive value-based purchasing}.
\newblock \bibinfo{journal}{Journal of Nursing Scholarship}
  \bibinfo{volume}{44}, \bibinfo{pages}{385--395}.
%Type = Article
\bibitem[{Kieschnick and McCullough(2003)}]{kieschnick2003regression}
\bibinfo{author}{Kieschnick, R.}, \bibinfo{author}{McCullough, B.D.},
  \bibinfo{year}{2003}.
\newblock \bibinfo{title}{Regression analysis of variates observed on (0, 1):
  percentages, proportions and fractions}.
\newblock \bibinfo{journal}{Statistical modelling} \bibinfo{volume}{3},
  \bibinfo{pages}{193--213}.
%Type = Article
\bibitem[{Kontopantelis et~al.(2015)Kontopantelis, Doran, Springate, Buchan and
  Reeves}]{kontopantelis2015regression}
\bibinfo{author}{Kontopantelis, E.}, \bibinfo{author}{Doran, T.},
  \bibinfo{author}{Springate, D.A.}, \bibinfo{author}{Buchan, I.},
  \bibinfo{author}{Reeves, D.}, \bibinfo{year}{2015}.
\newblock \bibinfo{title}{Regression based quasi-experimental approach when
  randomisation is not an option: interrupted time series analysis}.
\newblock \bibinfo{journal}{BMJ} \bibinfo{volume}{350}, \bibinfo{pages}{h2750}.
%Type = Article
\bibitem[{Masarotto and Varin(2012)}]{masarotto2012gaussian}
\bibinfo{author}{Masarotto, G.}, \bibinfo{author}{Varin, C.},
  \bibinfo{year}{2012}.
\newblock \bibinfo{title}{Gaussian copula marginal regression}.
\newblock \bibinfo{journal}{Electronic Journal of Statistics}
  \bibinfo{volume}{6}, \bibinfo{pages}{1517--1549}.
%Type = Book
\bibitem[{Nelsen(2007)}]{nelsen2007introduction}
\bibinfo{author}{Nelsen, R.B.}, \bibinfo{year}{2007}.
\newblock \bibinfo{title}{An introduction to copulas}.
\newblock \bibinfo{publisher}{Springer Science \& Business Media}.
%Type = Article
\bibitem[{Newey and West(1986)}]{newey1986simple}
\bibinfo{author}{Newey, W.K.}, \bibinfo{author}{West, K.D.},
  \bibinfo{year}{1986}.
\newblock \bibinfo{title}{A simple, positive semi-definite, heteroskedasticity
  and autocorrelation consistent covariance matrix}.
\newblock \bibinfo{journal}{Econometrica: journal of the Econometric Society}
  \bibinfo{volume}{55}, \bibinfo{pages}{703--708}.
%Type = Article
\bibitem[{Ospina and Ferrari(2012)}]{ospina2012general}
\bibinfo{author}{Ospina, R.}, \bibinfo{author}{Ferrari, S.L.},
  \bibinfo{year}{2012}.
\newblock \bibinfo{title}{A general class of zero-or-one inflated beta
  regression models}.
\newblock \bibinfo{journal}{Computational Statistics \& Data Analysis}
  \bibinfo{volume}{56}, \bibinfo{pages}{1609--1623}.
%Type = Article
\bibitem[{Penfold and Zhang(2013)}]{penfold2013use}
\bibinfo{author}{Penfold, R.B.}, \bibinfo{author}{Zhang, F.},
  \bibinfo{year}{2013}.
\newblock \bibinfo{title}{Use of interrupted time series analysis in evaluating
  health care quality improvements}.
\newblock \bibinfo{journal}{Academic pediatrics} \bibinfo{volume}{13},
  \bibinfo{pages}{S38--S44}.
%Type = Article
\bibitem[{Rhee et~al.(2021)Rhee, Wang, Ye, Baker, Griesbach, Laskowski and
  Klompas}]{rhee2021decline}
\bibinfo{author}{Rhee, C.}, \bibinfo{author}{Wang, R.}, \bibinfo{author}{Ye,
  S.}, \bibinfo{author}{Baker, M.A.}, \bibinfo{author}{Griesbach, D.},
  \bibinfo{author}{Laskowski, K.}, \bibinfo{author}{Klompas, M.},
  \bibinfo{year}{2021}.
\newblock \bibinfo{title}{Decline in sars-cov-2 infections among health care
  workers at 2 hospitals following rollout and administration of mrna
  vaccines}.
\newblock \bibinfo{journal}{Open Forum Infectious Diseases}
  \bibinfo{volume}{8}, \bibinfo{pages}{ofab204}.
%Type = Article
\bibitem[{S{\o}rensen(2019)}]{sorensen2019independence}
\bibinfo{author}{S{\o}rensen, H.}, \bibinfo{year}{2019}.
\newblock \bibinfo{title}{Independence, successive and conditional likelihood
  for time series of counts}.
\newblock \bibinfo{journal}{Journal of Statistical Planning and Inference}
  \bibinfo{volume}{200}, \bibinfo{pages}{20--31}.
%Type = Article
\bibitem[{Varin et~al.(2011)Varin, Reid and Firth}]{varin2011overview}
\bibinfo{author}{Varin, C.}, \bibinfo{author}{Reid, N.},
  \bibinfo{author}{Firth, D.}, \bibinfo{year}{2011}.
\newblock \bibinfo{title}{An overview of composite likelihood methods}.
\newblock \bibinfo{journal}{Statistica Sinica} , \bibinfo{pages}{5--42}.
%Type = Article
\bibitem[{Varin and Vidoni(2005)}]{varin2005note}
\bibinfo{author}{Varin, C.}, \bibinfo{author}{Vidoni, P.},
  \bibinfo{year}{2005}.
\newblock \bibinfo{title}{A note on composite likelihood inference and model
  selection}.
\newblock \bibinfo{journal}{Biometrika} \bibinfo{volume}{92},
  \bibinfo{pages}{519--528}.
%Type = Article
\bibitem[{Wagner et~al.(2002)Wagner, Soumerai, Zhang and
  Ross-Degnan}]{wagner2002segmented}
\bibinfo{author}{Wagner, A.K.}, \bibinfo{author}{Soumerai, S.B.},
  \bibinfo{author}{Zhang, F.}, \bibinfo{author}{Ross-Degnan, D.},
  \bibinfo{year}{2002}.
\newblock \bibinfo{title}{Segmented regression analysis of interrupted time
  series studies in medication use research}.
\newblock \bibinfo{journal}{Journal of clinical pharmacy and therapeutics}
  \bibinfo{volume}{27}, \bibinfo{pages}{299--309}.
%Type = Article
\bibitem[{Wang and Griswold(2017)}]{wang2017natural}
\bibinfo{author}{Wang, W.}, \bibinfo{author}{Griswold, M.E.},
  \bibinfo{year}{2017}.
\newblock \bibinfo{title}{Natural interpretations in tobit regression models
  using marginal estimation methods}.
\newblock \bibinfo{journal}{Statistical methods in medical research}
  \bibinfo{volume}{26}, \bibinfo{pages}{2622--2632}.
%Type = Article
\bibitem[{Ye et~al.(2022)Ye, Wang and Zhang}]{ye2022comparison}
\bibinfo{author}{Ye, S.}, \bibinfo{author}{Wang, R.}, \bibinfo{author}{Zhang,
  B.}, \bibinfo{year}{2022}.
\newblock \bibinfo{title}{Comparison of estimation methods and sample size
  calculation for parameter-driven interrupted time series models with count
  outcomes}.
\newblock \bibinfo{journal}{Health Services and Outcomes Research Methodology}
  , \bibinfo{pages}{1--48}.

\end{thebibliography}

\clearpage

\begin{table}[!h]
\centering
\caption{Bivariate copula functions.}
\begin{tabular}{c|l}
\hline
Copula & Copula function \\
\hline
Gaussian & $C(u_1, u_2; \rho) = \Phi_\rho (\Phi^{-1}(u_1), \Phi^{-1}(u_2))$, $\rho \in [-1, 1]$ \\
Clayton & $C(u_1, u_2; \rho) = \max(0, (u_1^{-\rho} + u_2^{-\rho} - 1)^{-1/\rho})$, $\rho \ge -1$ \\
Gumbel & $C(u_1, u_2; \rho) = \exp\{-[(-\log u_1)^\rho + (-\log u_1)^\rho]^{1/\rho} \}$, $\rho \ge 1$ \\
Frank & $C(u_1, u_2; \rho) = -\frac{1}{\rho} \log \left\{ 1 + \frac{[\exp(-\rho u_1) - 1] [\exp(-\rho u_2) - 1]}{\exp(-\rho - 1)} \right\}$, $\rho \in \mathbb{R}/\{0\}$ \\
Ali-Mikhail-Haq (AMH) & $C(u_1, u_2; \rho) = \frac{u_1 u_2}{1 - \rho (1 - u_1) (1 - u_2)}$, $\rho \in [-1, 1]$ \\
\hline
\end{tabular}
\label{Copula function}
\end{table}

\begin{table}[h]
\caption{Bias, standard error of $\hat{\beta}$ ($\mbox{SE}(\hat{\beta})$), mean of standard error estimates ($\E(\widehat{\mbox{SE}})$), 95\% confidence interval coverage probability (Cov.prob), and power (without model selection) for data generated from zero-one inflated Beta time series with Gaussian copula and fitted by Gaussian copula models.}
\label{Result1}
\begin{tabular}{c|c|c c c|c c c}
\hline
& & \multicolumn{3}{|c|}{Bootstrap}  & \multicolumn{3}{|c}{HAC} \\ \hline
& $\# \; n$ & 60 & 120 & 180& 60 & 120 & 180  \\ \hline
\multirow{10}{*}{$\rho=0.5$}&Bias $\beta_{32}$ &0.001 &-0.002&-0.003& -0.002&-0.006&-0.003\\ 
&SE($\hat{\beta}_{32}$) &0.272 & 0.155& 0.120& 0.157&0.132&0.115\\ 
&$E(\widehat{SE}_{\hat{\beta}_{32}})$ & 0.209 & 0.141& 0.121& 0.211& 0.143 &0.119 \\ 
&Cov.prob $\hat{\beta}_{32}$  & 0.966 & 0.943& 0.928& 0.830& 0.907 &0.924 \\ 
&Power $\hat{\beta}_{32}$  & 0.558 & 0.921& 0.983& 0.833& 0.955 &0.9895 \\ 

&Bias $\beta_{33}$ & -0.002 & 0.004& -0.004& -0.004& -0.008&-0.003 \\ 
&SE($\hat{\beta}_{33}$) &0.997 & 0.492& 0.277& 0.224&0.195&0.173 \\ 
&$E(\widehat{SE}_{\hat{\beta}_{33}})$ & 0.367 & 0.261&0.210&0.375&0.258&0.209\\ 
&Cov.prob $\hat{\beta}_{33}$& 0.934 & 0.926&0.931&0.735&0.847&0.874\\ 
&Power $\hat{\beta}_{33}$& 0.022 & 0.109&0.251&0.393&0.392&0.438\\ \hline

\multirow{10}{*}{$\rho=0.2$}&Bias $\beta_{32}$ &0.005 &0.002&-0.004& -0.011&-0.007&-0.001\\ 
&SE($\hat{\beta}_{32}$) &0.246 & 0.140& 0.109&0.147&0.122&0.105\\ 
&$E(\widehat{SE}_{\hat{\beta}_{32}})$ & 0.194 & 0.133& 0.108& 0.190& 0.136 &0.108 \\ 
&Cov.prob $\hat{\beta}_{32}$  & 0.964 & 0.944& 0.934&0.845&0.899&0.930 \\ 
&Power $\hat{\beta}_{33}$& 0.644 & 0.952&0.996&0.882&0.981&0.996\\ 

&Bias $\beta_{33}$ & 0.008 & 0.002& -0.002& -0.005& 0.008&-0.001 \\ 
&SE($\hat{\beta}_{33}$) &0.919 &0.411&0.239&0.210&0.178&0.157 \\ 
&$E(\widehat{SE}_{\hat{\beta}_{33}})$ &0.345& 0.239&0.195&0.345&0.242&0.190\\ 
&Cov.prob $\hat{\beta}_{33}$& 0.933 & 0.925&0.920&0.741&0.822&0.876\\
&Power $\hat{\beta}_{33}$&0.023 & 0.158&0.300&0.406&0.428&0.487\\
\hline

\multirow{10}{*}{$\rho=-0.2$}&Bias $\beta_{32}$ &0.003 &0.001&0.003& -0.009&-0.001&-0.001\\ 
&SE($\hat{\beta}_{32}$) &0.210 & 0.117& 0.091& 0.132&0.106&0.092\\ 
&$E(\widehat{SE}_{\hat{\beta}_{32}})$ & 0.166 & 0.112& 0.093& 0.162& 0.116 &0.093 \\ 
&$Cov.prob \; \hat{\beta}_{32}$  & 0.966 & 0.944& 0.927& 0.851& 0.907 &0.936 \\ 
&$Power \; \hat{\beta}_{32}$  & 0.763 & 0.991& 1.000& 0.944& 0.996 &1.000 \\ 

&Bias $\beta_{33}$ & -0.002 & -0.005& -0.006& 0.011& 0.003&0.000 \\ 
&SE($\hat{\beta}_{33}$) &0.787 & 0.345& 0.187& 0.184&0.155&0.136 \\ 
&$E(\widehat{SE}_{\hat{\beta}_{33}})$ & 0.300 & 0.205&0.164&0.292&0.203&0.160\\ 
&$Cov.prob \; \hat{\beta}_{33}$& 0.941 & 0.927&0.926&0.755&0.839&0.888\\
&$Power \; \hat{\beta}_{33}$  & 0.046 & 0.225& 0.431& 0.429& 0.490 &0.591 \\ \hline

\multirow{10}{*}{$\rho=-0.5$}&Bias $\beta_{32}$ &0.003 &-0.002&-0.002& -0.003&0.001&-0.001\\ 
&SE($\hat{\beta}_{32}$) &0.198 & 0.107& 0.082& 0.122&0.095&0.081\\ 
&$E(\widehat{SE}_{\hat{\beta}_{32}})$ & 0.144 & 0.098& 0.080& 0.152& 0.099 &0.079 \\ 
&$Cov.prob \; \hat{\beta}_{32}$  & 0.969 & 0.956& 0.935& 0.857& 0.925 &0.939 \\ 
&$Power \; \hat{\beta}_{32}$  & 0.802 & 0.996& 1.000& 0.954& 0.999 &1.000 \\ 

&Bias $\beta_{33}$ & 0.006 & 0.003& 0.004& -0.009& 0.000&0.001 \\ 
&SE($\hat{\beta}_{33}$) &0.742 & 0.311& 0.173& 0.170&0.138&0.121 \\ 
&$E(\widehat{SE}_{\hat{\beta}_{33}})$ & 0.265 & 0.167&0.139&0.255&0.172&0.139\\ 
&$Coverage \; \hat{\beta}_{33}$& 0.942 & 0.945&0.934&0.793&0.867&0.903\\ 
&$Power \; \hat{\beta}_{33}$  & 0.038 & 0.254& 0.508& 0.495& 0.574 &0.676 \\ \hline
\end{tabular}
\end{table}

\begin{table}[h]
\caption{Bias, standard error of $\hat{\beta}$ ($\mbox{SE}(\hat{\beta})$), mean of standard error estimates ($\E(\widehat{\mbox{SE}})$), 95\% confidence interval coverage probability (Cov.prob), and power (without model selection) for data generated from zero-one inflated Beta time series with Frank copula and fitted by Gaussian copula models.}
\label{Result2}
\begin{tabular}{c|c|c c c|c c c}
\hline
& & \multicolumn{3}{|c|}{Bootstrap}  & \multicolumn{3}{|c}{HAC} \\ \hline
& $\# \; n$ & 60 & 120 & 180& 60 & 120 & 180  \\ \hline
\multirow{10}{*}{$\rho=3.3$}&Bias $\beta_{32}$ &0.008 &-0.007&-0.001& 0.010&-0.002&-0.001\\ 
&SE($\hat{\beta}_{32}$) &0.281 & 0.155& 0.120& 0.164&0.133&0.116\\ 
&$E(\widehat{SE}_{\hat{\beta}_{32}})$ & 0.209 & 0.148& 0.123& 0.209& 0.150 &0.124 \\ 
&Cov.prob $\hat{\beta}_{32}$  & 0.973 & 0.949& 0.922& 0.845& 0.889 &0.918 \\ 
&Power $\hat{\beta}_{32}$  & 0.523 & 0.925& 0.984& 0.808& 0.952 &0.987 \\ 

&Bias $\beta_{33}$ & 0.011 & 0.009& -0.002& -0.014& -0.001&0.002 \\ 
&SE($\hat{\beta}_{33}$) &1.069 & 0.480& 0.265& 0.230&0.197&0.175 \\ 
&$E(\widehat{SE}_{\hat{\beta}_{33}})$ & 0.382 & 0.265&0.218&0.387&0.267&0.217\\ 
&Cov.prob $\hat{\beta}_{33}$& 0.931 & 0.918&0.923&0.738&0.825&0.869\\ 
&Power $\hat{\beta}_{33}$& 0.009 & 0.133&0.275&0.410&0.377&0.434\\ \hline

\multirow{10}{*}{$\rho=1.17$}&Bias $\beta_{32}$ &0.013 &-0.003&0.000& 0.004&-0.004&0.000\\ 
&SE($\hat{\beta}_{32}$) &0.246 & 0.140& 0.108&0.151&0.123&0.106\\ 
&$E(\widehat{SE}_{\hat{\beta}_{32}})$ & 0.193 & 0.136& 0.110& 0.194& 0.135 &0.112 \\ 
&Cov.prob $\hat{\beta}_{32}$  & 0.968 & 0.943& 0.929&0.834&0.902&0.925 \\ 
&Power $\hat{\beta}_{33}$& 0.625 & 0.954&0.992&0.866&0.976&0.997\\ 

&Bias $\beta_{33}$ & 0.001 & -0.004& -0.001& 0.004& 0.002&0.009 \\ 
&SE($\hat{\beta}_{33}$) &0.937 &0.404&0.230&0.201&0.181&0.160 \\ 
&$E(\widehat{SE}_{\hat{\beta}_{33}})$ &0.352& 0.247&0.198&0.339&0.243&0.195\\ 
&Cov.prob $\hat{\beta}_{33}$& 0.929 & 0.918&0.922&0.730&0.837&0.876\\
&Power $\hat{\beta}_{33}$&0.021 & 0.160&0.324&0.410&0.436&0.461\\
\hline

\multirow{10}{*}{$\rho=-1.17$}&Bias $\beta_{32}$ &0.006 &-0.004&0.000& 0.003&-0.002&-0.001\\ 
&SE($\hat{\beta}_{32}$) &0.215 & 0.118& 0.092& 0.130&0.104&0.089\\ 
&$E(\widehat{SE}_{\hat{\beta}_{32}})$ & 0.163 & 0.112& 0.089& 0.115& 0.091 &0.093 \\ 
&$Cov.prob \; \hat{\beta}_{32}$  & 0.961 & 0.944& 0.934& 0.850& 0.903 &0.929 \\ 
&$Power \; \hat{\beta}_{32}$  & 0.754 & 0.991& 1.000& 0.942& 0.993 &1.000 \\ 

&Bias $\beta_{33}$ & -0.001 & 0.002& -0.003& 0.010& 0.002&-0.003 \\ 
&SE($\hat{\beta}_{33}$) &0.820 & 0.343& 0.191& 0.180&0.152&0.132 \\ 
&$E(\widehat{SE}_{\hat{\beta}_{33}})$ & 0.280 & 0.198&0.160&0.289&0.197&0.157\\ 
&$Cov.prob \; \hat{\beta}_{33}$& 0.956 & 0.933&0.936&0.760&0.851&0.879\\
&$Power \; \hat{\beta}_{33}$  & 0.034 & 0.213& 0.437& 0.448& 0.510 &0.617 \\ \hline

\multirow{10}{*}{$\rho=-3.3$}&Bias $\beta_{32}$ &0.008 &-0.002&0.001&0.000&-0.001&-0.001\\ 
&SE($\hat{\beta}_{32}$) &0.198 & 0.107& 0.082& 0.122&0.095&0.081\\ 
&$E(\widehat{SE}_{\hat{\beta}_{32}})$ & 0.140 & 0.093& 0.077& 0.140& 0.095 &0.076 \\ 
&$Cov.prob \; \hat{\beta}_{32}$  & 0.971 & 0.963& 0.954& 0.891& 0.932 &0.942 \\ 
&$Power \; \hat{\beta}_{32}$  & 0.807 & 0.998& 1.000& 0.963& 0.999 &1.000 \\ 

&Bias $\beta_{33}$ & 0.004 & -0.005& -0.002& -0.003& -0.006&0.000 \\ 
&SE($\hat{\beta}_{33}$) &0.765 & 0.321& 0.172& 0.167&0.134&0.116 \\ 
&$E(\widehat{SE}_{\hat{\beta}_{33}})$ & 0.253 & 0.168&0.131&0.250&0.164&0.134\\ 
&$Coverage \; \hat{\beta}_{33}$& 0.952 & 0.951&0.949&0.801&0.879&0.892\\ 
&$Power \; \hat{\beta}_{33}$  & 0.037 & 0.270& 0.519& 0.495& 0.614 &0.704 \\ \hline
\end{tabular}
\end{table}

\begin{table}[h]
\caption{Bias, standard error of $\hat{\beta}$ ($\mbox{SE}(\hat{\beta})$), mean of standard error estimates ($\E(\widehat{\mbox{SE}})$), 95\% confidence interval coverage probability (Cov.prob), and power (with model selection) for data generated from zero-one inflated Beta time series with Gaussian copula and fitted by Gaussian copula models.}
\label{Result3}
\begin{tabular}{c|c|c c c|c c c}
\hline
& & \multicolumn{3}{|c|}{Bootstrap}  & \multicolumn{3}{|c}{HAC} \\ \hline
& $\# \; n$ & 60 & 120 & 180& 60 & 120 & 180  \\ \hline
\multirow{10}{*}{$\rho=0.5$}&Bias $\beta_{32}$ &-0.027 &-0.020& -0.016 & -0.033&-0.022&-0.014\\ 
&SE($\hat{\beta}_{32}$) &0.189 & 0.140& 0.116& 0.158&0.131&0.114\\ 
&$E(\widehat{SE}_{\hat{\beta}_{32}})$ & 0.223 & 0.146& 0.120& 0.222& 0.148 &0.124 \\ 
&Cov.prob $\hat{\beta}_{32}$  & 0.904 & 0.932& 0.930& 0.826& 0.891 &0.905 \\ 
&Power $\hat{\beta}_{32}$  & 0.786 & 0.951& 0.992& 0.872& 0.967 &0.993 \\ 

&Bias $\beta_{33}$ & 0.025 & 0.022& 0.011& 0.043& 0.028&0.010 \\ 
&SE($\hat{\beta}_{33}$) &0.334 & 0.246& 0.203& 0.213&0.188&0.168 \\ 
&$E(\widehat{SE}_{\hat{\beta}_{33}})$ & 0.399 & 0.270&0.217&0.406&0.275&0.217\\ 
&Cov.prob $\hat{\beta}_{33}$& 0.886 & 0.914&0.922&0.668&0.796&0.850\\ 
&Power $\hat{\beta}_{33}$& 0.197 & 0.226&0.310&0.436&0.386&0.444\\ \hline

\multirow{10}{*}{$\rho=0.2$}&Bias $\beta_{32}$ &-0.025&-0.017&-0.015& -0.034&-0.016&-0.009\\ 
&SE($\hat{\beta}_{32}$) &0.176 & 0.130& 0.107&0.144&0.120&0.104\\ 
&$E(\widehat{SE}_{\hat{\beta}_{32}})$ & 0.201 & 0.134& 0.107& 0.203& 0.136 &0.108 \\ 
&Cov.prob $\hat{\beta}_{32}$  & 0.900 & 0.936& 0.946&0.826&0.895&0.928 \\ 
&Power $\hat{\beta}_{33}$& 0.835 & 0.972&0.998&0.919&0.979&0.998\\ 

&Bias $\beta_{33}$ & 0.034 & 0.010& 0.017& 0.048& 0.010&0.002 \\ 
&SE($\hat{\beta}_{33}$) &0.310 &0.228&0.188&0.197&0.172&0.153 \\ 
&$E(\widehat{SE}_{\hat{\beta}_{33}})$ &0.368& 0.256&0.197&0.379&0.246&0.195\\ 
&Cov.prob $\hat{\beta}_{33}$& 0.889 & 0.909&0.932&0.679&0.811&0.854\\
&Power $\hat{\beta}_{33}$&0.202 & 0.285&0.342&0.443&0.434&0.492\\
\hline

\multirow{10}{*}{$\rho=-0.2$}&Bias $\beta_{32}$ &-0.031 &-0.019&-0.010& -0.024&-0.016&-0.010\\ 
&SE($\hat{\beta}_{32}$) &0.156 & 0.111& 0.090& 0.130&0.104&0.091\\ 
&$E(\widehat{SE}_{\hat{\beta}_{32}})$ & 0.171 & 0.112& 0.091& 0.172& 0.115 &0.093 \\ 
&$Cov.prob \; \hat{\beta}_{32}$  & 0.918 & 0.940& 0.940& 0.853& 0.906 &0.928 \\ 
&$Power \; \hat{\beta}_{32}$  & 0.910 & 0.997& 1.000& 0.943& 0.995 &1.000 \\ 

&Bias $\beta_{33}$ & 0.025 & 0.014& 0.004& 0.034& 0.017&0.008 \\ 
&SE($\hat{\beta}_{33}$) &0.272 & 0.194& 0.157& 0.173&0.151&0.135 \\ 
&$E(\widehat{SE}_{\hat{\beta}_{33}})$ & 0.308 & 0.209&0.163&0.306&0.208&0.167\\ 
&$Cov.prob \; \hat{\beta}_{33}$& 0.911 & 0.926&0.939&0.723&0.822&0.869\\
&$Power \; \hat{\beta}_{33}$  & 0.226 & 0.345& 0.475& 0.456& 0.484 &0.572 \\ \hline

\multirow{10}{*}{$\rho=-0.5$}&Bias $\beta_{32}$ &-0.021&-0.015&-0.012& -0.025&-0.011&-0.007\\ 
&SE($\hat{\beta}_{32}$) &0.145 & 0.100& 0.081& 0.121&0.096&0.081\\ 
&$E(\widehat{SE}_{\hat{\beta}_{32}})$ & 0.147 & 0.097& 0.079& 0.152& 0.100 &0.080 \\ 
&$Cov.prob \; \hat{\beta}_{32}$  & 0.945 & 0.956& 0.957& 0.866& 0.924 &0.946 \\ 
&$Power \; \hat{\beta}_{32}$  & 0.939 & 0.999& 1.000& 0.962& 0.999 &1.000 \\ 

&Bias $\beta_{33}$ & 0.031 & 0.012& 0.011& 0.021& 0.013&0.001 \\ 
&SE($\hat{\beta}_{33}$) &0.251 & 0.173& 0.140& 0.164&0.136&0.120 \\ 
&$E(\widehat{SE}_{\hat{\beta}_{33}})$ & 0.283 & 0.180&0.143&0.279&0.180&0.140\\ 
&$Coverage \; \hat{\beta}_{33}$& 0.910 & 0.935&0.944&0.741&0.852&0.898\\ 
&$Power \; \hat{\beta}_{33}$  & 0.246 & 0.396& 0.537& 0.479& 0.562 &0.694 \\ \hline
\end{tabular}
\end{table}

\begin{table}[h]
\caption{Bias, standard error of $\hat{\beta}$ ($\mbox{SE}(\hat{\beta})$), mean of standard error estimates ($\E(\widehat{\mbox{SE}})$), 95\% confidence interval coverage probability (Cov.prob), and power (with model selection) for data generated from zero-one inflated Beta time series with Frank copula and fitted by Gaussian copula models.}
\label{Result4}
\begin{tabular}{c|c|c c c|c c c}
\hline
& & \multicolumn{3}{|c|}{Bootstrap}  & \multicolumn{3}{|c}{HAC} \\ \hline
& $\# \; n$ & 60 & 120 & 180& 60 & 120 & 180  \\ \hline
\multirow{10}{*}{$\rho=3.3$}&Bias $\beta_{32}$ &0.003 &-0.002&-0.002& -0.003&0.001&-0.001\\ 
&SE($\hat{\beta}_{32}$) &0.198 & 0.107& 0.082& 0.122&0.096&0.081\\ 
&$E(\widehat{SE}_{\hat{\beta}_{32}})$ & 0.145 & 0.098& 0.080& 0.152& 0.099 &0.079 \\ 
&Cov.prob $\hat{\beta}_{32}$  & 0.969 & 0.956& 0.936& 0.857& 0.925 &0.939 \\ 
&Power $\hat{\beta}_{32}$  & 0.802 & 0.996& 1.000& 0.954& 0.999 &1.000 \\ 

&Bias $\beta_{33}$ & 0.006 & 0.003& 0.004& -0.009& 0.000&0.001 \\ 
&SE($\hat{\beta}_{33}$) &0.742 & 0.311& 0.173& 0.170&0.138&0.121 \\ 
&$E(\widehat{SE}_{\hat{\beta}_{33}})$ & 0.265& 0.167&0.139&0.255&0.172&0.139\\ 
&Cov.prob $\hat{\beta}_{33}$& 0.942 & 0.945&0.934&0.793&0.867&0.903\\ 
&Power $\hat{\beta}_{33}$& 0.038 & 0.254&0.508&0.495&0.574&0.676\\ \hline

\multirow{10}{*}{$\rho=1.17$}&Bias $\beta_{32}$ &0.008 &-0.002&0.001& 0.000&-0.001&-0.001\\ 
&SE($\hat{\beta}_{32}$) &0.198 & 0.106& 0.083&0.122&0.092&0.078\\ 
&$E(\widehat{SE}_{\hat{\beta}_{32}})$ & 0.140 & 0.093& 0.077& 0.140& 0.095 &0.076 \\ 
&Cov.prob $\hat{\beta}_{32}$  & 0.971 & 0.963& 0.954&0.891&0.932&0.942 \\ 
&Power $\hat{\beta}_{33}$& 0.807 & 0.998&1.000&0.963&0.999&1.000\\ 

&Bias $\beta_{33}$ & 0.004 & -0.005& -0.002& -0.003& -0.006&0.000 \\ 
&SE($\hat{\beta}_{33}$) &0.765 &0.321&0.172&0.167&0.134&0.116 \\ 
&$E(\widehat{SE}_{\hat{\beta}_{33}})$ &0.253& 0.168&0.131&0.250&0.164&0.134\\ 
&Cov.prob $\hat{\beta}_{33}$& 0.952 & 0.951&0.949&0.801&0.879&0.892\\
&Power $\hat{\beta}_{33}$&0.037 &0.270&0.519& 0.495&0.614&0.704\\
\hline

\multirow{10}{*}{$\rho=-1.17$}&Bias $\beta_{32}$ &-0.021 &-0.015&-0.012& -0.025&-0.011&-0.007\\ 
&SE($\hat{\beta}_{32}$) &0.145 & 0.100& 0.081& 0.121&0.096&0.081\\ 
&$E(\widehat{SE}_{\hat{\beta}_{32}})$ & 0.147 & 0.097& 0.079& 0.152& 0.101 &0.080 \\ 
&$Cov.prob \; \hat{\beta}_{32}$  & 0.945 & 0.956& 0.957& 0.866& 0.924 &0.946 \\ 
&$Power \; \hat{\beta}_{32}$  & 0.939 & 0.999& 1.000& 0.962& 0.999 &1.000 \\ 

&Bias $\beta_{33}$ & 0.031 & 0.012& 0.011& 0.021& 0.013&0.001 \\ 
&SE($\hat{\beta}_{33}$) &0.251 & 0.173& 0.140& 0.164&0.136&0.120 \\ 
&$E(\widehat{SE}_{\hat{\beta}_{33}})$ & 0.283 & 0.180&0.143&0.279&0.180&0.140\\ 
&$Cov.prob \; \hat{\beta}_{33}$& 0.910 & 0.935&0.944&0.741&0.852&0.898\\
&$Power \; \hat{\beta}_{33}$  & 0.246 & 0.396& 0.537& 0.479& 0.562 &0.694 \\ \hline

\multirow{10}{*}{$\rho=-3.3$}&Bias $\beta_{32}$ &-0.024 &-0.015&-0.008&-0.020&-0.014&-0.006\\ 
&SE($\hat{\beta}_{32}$) &0.147 & 0.101& 0.082& 0.117&0.092&0.077\\ 
&$E(\widehat{SE}_{\hat{\beta}_{32}})$ & 0.144 & 0.093& 0.074& 0.138& 0.094 &0.079 \\ 
&$Cov.prob \; \hat{\beta}_{32}$  & 0.952 & 0.962& 0.968& 0.877& 0.940 &0.939 \\ 
&$Power \; \hat{\beta}_{32}$  & 0.951 & 0.999& 1.000& 0.981& 0.999 &1.000 \\ 

&Bias $\beta_{33}$ & 0.022 & 0.006& 0.001& 0.018& 0.011&0.005 \\ 
&SE($\hat{\beta}_{33}$) &0.254 & 0.174& 0.141& 0.159&0.131&0.115 \\ 
&$E(\widehat{SE}_{\hat{\beta}_{33}})$ & 0.268 & 0.166&0.135&0.265&0.172&0.135\\ 
&$Coverage \; \hat{\beta}_{33}$& 0.933 & 0.959&0.948&0.754&0.865&0.896\\ 
&$Power \; \hat{\beta}_{33}$  & 0.227 & 0.407& 0.551& 0.508& 0.604 &0.714 \\ \hline
\end{tabular}
\end{table}

\begin{table}[!h]
\centering
\caption{The 95\% confidence intervals, $p$-values, and estimates of the mean parameters.}
\begin{tabular}{c|c|c|c}
\hline
Parameters & Point estimates & 95\% Confidence Intervals & $p$-values  \\
\hline
$\beta_{30}$: initial level & $0.512$ & $(0.075, 0.948)$ & \bf 0.021 \\
$\beta_{31}$: initial trend & $0.018$ & $(-0.005, 0.041)$ & 0.119 \\
$\beta_{32}$: level change & $0.362$ & $(-0.275, 1.000)$ & 0.265 \\
$\beta_{33}$: trend change & $-0.029$ & $(-0.064, 0.006)$ & 0.103 \\
$\beta_{40}$: initial dispersion & $2.196$ & $(1.715, 2.677)$ & \bf 0.000 \\
$\beta_{41}$: dispersion change & $0.795$ & $(0.104, 1.486)$ & \bf 0.024 \\
\hline
\end{tabular}
\label{App Results}
\end{table}

\begin{figure}[h]
    \centering
    \includegraphics[width=1\linewidth]{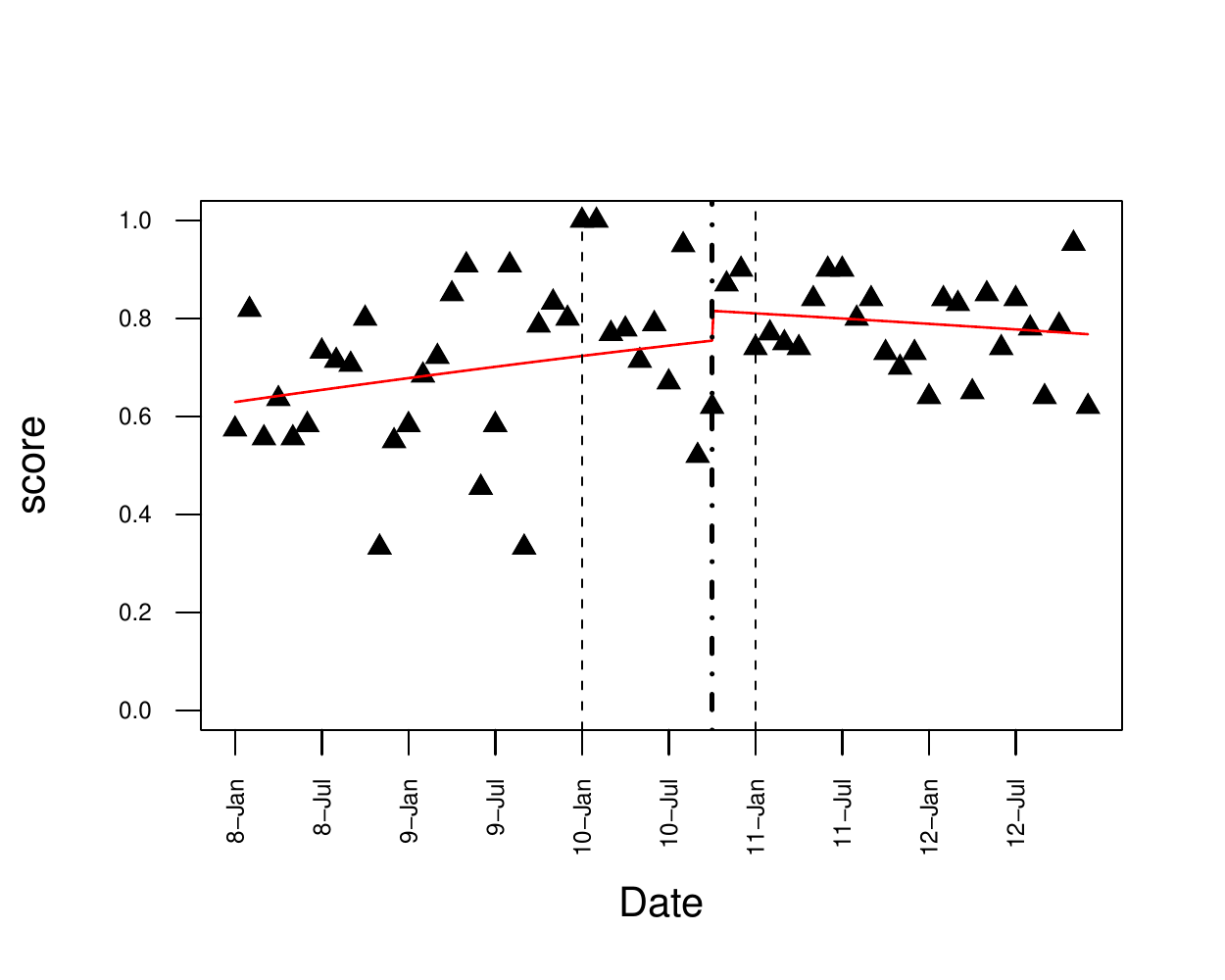}
    \caption{The time series of observed `pain management' scores for a single hospital, the estimated change point, estimated means, and boundaries of possible change points. The estimated means and the change point are obtained from modeling the time series with MZOIBTS.}
    \label{Data}
\end{figure}

\end{document}